\documentclass[12pt,amsmath,amssymb]{revtex4}
\usepackage{amsmath,amsfonts,amsthm,amscd,amssymb,latexsym}
\usepackage{bm}
\usepackage{dcolumn}
\usepackage{graphicx}
\usepackage{epstopdf}
\usepackage{color}
\usepackage{epsf}
\usepackage{epsfig}
\usepackage{graphicx, epic, eepic, color}

\begin{document}

\title{Quasinormal Modes of a Black Hole with Quadrupole Moment}

\author{Alireza \surname{Allahyari}$^{1}$}
\email{alireza.al@ipm.ir}
\author{Hassan \surname{Firouzjahi}$^{1}$}
\email{firouz@ipm.ir}
\author{Bahram \surname{Mashhoon}$^{1,2}$}
\email{mashhoonb@missouri.edu}

\affiliation{$^1$School of Astronomy, Institute for Research in Fundamental
Sciences (IPM), P. O. Box 19395-5531, Tehran, Iran\\
$^2$Department of Physics and Astronomy, University of Missouri, Columbia, Missouri 65211, USA\\
}

\date{\today}

\begin{abstract}
We analytically determine the quasinormal mode (QNM) frequencies of a black hole with quadrupole moment in the eikonal limit using the light-ring method.  The generalized black holes that are discussed in this work possess arbitrary quadrupole and higher mass moments in addition to mass and angular momentum.  Static collapsed configurations with mass and quadrupole moment are treated in detail and the QNM frequencies associated with two such configurations are evaluated to linear order in the quadrupole moment.  Furthermore, we touch upon the treatment of rotating systems. In particular, the generalized black hole that we consider for our extensive QNM calculations is a completely collapsed configuration whose exterior gravitational field can be described by the Hartle-Thorne spacetime [Astrophys. J. {\bf 153}, 807-834 (1968)]. This collapsed system as well as its QNMs is characterized by mass $M$, quadrupole moment $Q$ and angular momentum $J$, where the latter two parameters are treated to first and second orders of approximation, respectively. When the quadrupole moment is set equal to the relativistic quadrupole moment of the corresponding Kerr black hole, $J^2/(Mc^2)$, the Hartle-Thorne QNMs reduce to those of the Kerr black hole to second order in angular momentum $J$. Using ringdown frequencies, one cannot observationally distinguish a generalized Hartle-Thorne black hole with arbitrary quadrupole moment from a Kerr black hole provided the dimensionless parameter given by $|QMc^2-J^2|c^2/(G^2M^4)$ is sufficiently small compared to unity. 
\end{abstract}

\maketitle

\section{Introduction}

To date, ten detections of gravitational waves have been interpreted in terms of binary black hole mergers~\cite{Abbott:2016blz, Abbott:2016nmj, TheLIGOScientific:2016pea, Abbott:2017gyy, Abbott:2017vtc, Abbott:2017oio, NEW}. The reported identifications are based on the gravitational ringdown of the final collapsed configuration as it settles into a relatively quiescent state. Future observations of gravitational waves are expected to probe the complete gravitational collapse of matter. Black holes are Ricci-flat solutions of the gravitational field equations in general relativity that represent the complete gravitational collapse of matter to singular states, where such singularities are hidden from the outside world via the existence of regular event horizons~\cite{Chandra}. However, more general collapsed configurations are possible within the classical framework of general relativity. 

Consider the complete gravitational collapse of an astronomical system in accordance with classical general relativity. We expect that the final state of collapse is characterized by a set of exterior  multipole moments of the system. Among such final states, isolated black holes form a very special stationary axisymmetric subclass in which the intrinsic Newtonian quadrupole moments as well as the higher nonrelativistic moments of the original systems have all been radiated away in the collapse process. In standard general relativity, a black hole final state is given in general  by the charged Kerr spacetime characterized by  its mass $M$, electric charge $e_{BH}$ and angular momentum $J$. The exterior Kerr spacetime does have a \emph{relativistic} quadrupole moment given by $J^2/(Mc^2)$, which vanishes in the Newtonian limit ($c \to \infty$).

The remaining possible final states would have a more complex structure in comparison with black holes. Indeed, black hole uniqueness theorems imply that one should be dealing with naked singularities in such cases, since the resulting spacetime singularities would not be covered by event horizons.  The final states may contain all gravitoelectric and gravitomagnetic multipole moments. In the space of such final states, black holes occupy a set of measure zero.  The temporal decay of higher moments during the collapse process has been investigated within the framework of linear perturbation theory by Price~\cite{Price:1971fb}; however, a general nonlinear treatment does not exist. The formation of a regular event horizon has the advantage that the exterior region satisfies the principle of causality. To preserve this, Penrose has conjectured that singularities that would actually occur in nature must be black holes (Cosmic Censorship Conjecture). However, there is no general proof of this conjecture and most singularities that are theoretically encountered  in general relativity are naked singularities, such as the Big Bang singularity~\cite{Joshi:2015uoq}. General relativity breaks down at a spacetime singularity; therefore, the actual state of gravitationally collapsed matter is at present unknown. 

In this paper, we consider the possibility that the collapsed configurations involved in the gravitational wave observations might in fact be \emph{generalized black holes} possessing arbitrary multipole moments. A class of Ricci-flat stationary axisymmetric spacetimes that is a generalization of Kerr spacetime with all multipole moments is described in the next section.   Is the theory associated with black hole quasinormal modes (QNMs) applicable to generalized black holes? The corresponding perturbation equations do not appear separable and it is not known whether such systems can undergo regular quasinormal mode oscillations. We address this issue in the present paper; moreover,  in this first treatment of the oscillations of such systems, we limit our considerations to the simplest generalized collapsed configuration, namely, a black hole of mass $M$ possessing angular momentum $J$ and a classical quadrupole moment $Q$. To simplify matters even further, we treat the angular momentum to second order and the quadrupole contribution to first order.  We study analytically the QNMs of this configuration in the eikonal limit using the light-ring method. 

The plan of this paper is as follows. Generalized Kerr spacetimes are briefly introduced in Section II. There is no known uniqueness result associated with such gravitational fields; indeed, there are other known classes of solutions of this type. We study exact solutions representing collapsed systems with mass and quadrupole moment in Section III for the sake of simplicity. Further simplification occurs when we focus on linearized quadrupolar perturbations of a Schwarzschild black hole. In fact, two such solutions are considered in some detail in Section IV; that is, we introduce the $SQ$ solution and discuss the static Hartle-Thorne ($HT$) solution. The analytic light-ring method is described in Section V and the QNM frequencies of the $SQ$ spacetime and the static $HT$ spacetime are calculated in the eikonal limit on the basis of the temporal decay of a congruence of equatorial null rays away from the unstable light ring. It is noteworthy that we find essentially the \emph{same} QNM frequencies. To include rotation in our treatment, we choose the Hartle-Thorne solution due to its physical significance. Section VI generalizes our result to the QNM frequencies of the rotating Hartle-Thorne solution. A discussion of our results is contained in Section VII. Some calculations are relegated to the appendices. Throughout this paper, Greek indices run from $0$ to $3$, while Latin indices run from $1$ to $3$. The signature of the spacetime metric is $+2$ and units are chosen such that $c = G = 1$, unless specified otherwise.

\section{Generalized Kerr Spacetimes}

The static exterior gravitational field of a nonrotating axisymmetric body was first described by Weyl in 1917~\cite{Weyl}. In terms of Weyl's canonical (cylindrical) coordinates  $(t, \rho, \phi, z)$, the line element for this field can be expressed as 
\begin{equation}\label{eq: 1}
ds^2=-e^{2 \psi} \,dt^2 +e^{-2\psi}\,[e^{2\gamma}(d\rho^2+dz^2)+\rho^2d\phi^2]\,, 
\end{equation}                                     
where the metric functions $\psi $       and $\gamma$     are functions of  $\rho$      and $z$, and
\begin{equation}\label{eq: 2}
\psi,_{\rho\rho}+\frac{1}{\rho}\,\psi,_\rho + \psi,_{zz}=0\,,
\end{equation}
\begin{equation}\label{eq: 3}
\gamma,_\rho=\rho\,(\psi,_\rho^2 - \psi,_z^2)\,, \qquad \gamma,_z=2\,\rho\, \psi,_\rho\psi,_z\,.
\end{equation}
Here, a comma denotes partial differentiation, so that $\psi,_\rho := \partial \psi/ \partial \rho$.
Equation~\eqref{eq: 2} is the integrability condition for the relations in Eq.~\eqref{eq: 3}; that is, $\gamma_{,\rho z} = \gamma_{,z \rho}$, once $\psi$ is a solution of Eq.~\eqref{eq: 2}. Moreover, it is a remarkable fact that Eq.~\eqref{eq: 2} is Laplace's equation in cylindrical coordinates; that is, Eq.~\eqref{eq: 2} is equivalent to $\nabla^2\psi=0$ and $\psi$ is therefore a harmonic function. It is natural to associate $\psi$ with the exterior Newtonian gravitational potential of a source, which in general has nothing to do with the source of the exterior general relativistic gravitational field. There are two commuting hypersurface-orthogonal Killing vector fields $\partial_t$ and $\partial_\phi$ in this static axisymmetric system. Furthermore, the symmetry axis is regular (i.e., elementary flat) if $\gamma(\rho, z)$    vanishes  as $\rho\to0$. 

If the gravitational potentials $\psi$ and $\gamma$ vanish, then metric~\eqref{eq: 1} represents Minkowski spacetime in standard spatial cylindrical coordinate system.  An important feature of the field Eqs.~\eqref{eq: 2}--\eqref{eq: 3} should be noted here:  From a given solution $(\psi, \gamma)$, one can generate a set of solutions $(\delta\,\psi, \delta^2\,\gamma)$, where $\delta$ is a real nonzero parameter. This $\delta$-transformation, namely, $(\psi, \gamma) \mapsto (\delta\,\psi, \delta^2\,\gamma)$, plays an important role in the considerations of this paper. For background material regarding various aspects of exact solutions representing the exterior gravitational field of static and stationary axisymmetric configurations in general relativity, we refer to Refs.~\cite{Syng, SKMH, Griffiths:2009dfa}.

It proves useful to introduce prolate spheroidal coordinates $(t, x, y, \phi)$ that are related to the Weyl canonical coordinates  $(t, \rho, \phi, z)$ via
\begin{equation}\label{eq: 4}
x=\frac{1}{2\sigma}(r_+ + r_-)\,,  \qquad             y=\frac{1}{2\sigma}(r_+ - r_-)\,, \qquad r_\pm^2=\rho^2+(z\pm\sigma)^2,
\end{equation}
where  $x\ge 1$,      $-1\le y \le 1$ and   $\sigma > 0$ is a constant length.
Let us now transform Weyl's metric to prolate spheroidal coordinates $(t, x, y, \phi)$. The result is
\begin{equation}\label{eq: 5}
ds^2=-e^{2 \psi} \,dt^2 + \sigma^2\,e^{2\gamma-2\psi}\,(x^2-y^2)\left(\frac{dx^2}{x^2-1} + \frac{dy^2}{1-y^2}\right) + \sigma^2 e^{-2\psi} (x^2-1)(1-y^2)\,d\phi^2\,, 
\end{equation}   
where $\psi(x, y)$ and $\gamma(x, y)$ are now solutions of
\begin{equation}\label{eq: 6}
[(x^2-1)\,\psi,_x],_x + [(1-y^2)\,\psi,_y],_y=0\,,
\end{equation}
\begin{equation}\label{eq: 7}
\gamma,_x = \frac{1-y^2}{x^2-y^2}\,[x(x^2-1)\,\psi,_x^2 - x(1-y^2)\,\psi,_y^2 -2y (x^2-1)\,\psi,_x \psi,_y]\,,
\end{equation}
\begin{equation}\label{eq: 8}
\gamma,_y = \frac{x^2-1}{x^2-y^2}\,[y(x^2-1)\,\psi,_x^2 - y(1-y^2)\,\psi,_y^2 + 2x (1-y^2)\,\psi,_x \psi,_y]\,.
\end{equation}
In these equations, the requirement of asymptotic flatness means that as $x \to \infty$, we have $\psi(x, y) \to 0$ and $\gamma(x, y) \to 0$. Moreover, elementary flatness requires that $\gamma(x, \pm1) = 0$.

We now look for a solution of Laplace's equation in prolate spheroidal coordinates  via separation of variables, namely, 
\begin{equation}\label{eq: 9}
\psi(x, y) = A(x) B(y)\,.
\end{equation}
Substituting this ansatz in Eq.~\eqref{eq: 6}, we find the same differential equation for both $A(x)$ and $B(y)$; that is, 
\begin{equation}\label{eq: 10}
\frac{d}{dx}\left[(x^2-1)\,\frac{dA}{dx}\right] - \lambda_0\, A(x) = 0\,, \qquad  \frac{d}{dy}\left[(1-y^2)\,\frac{dB}{dy}\right] + \lambda_0\, B(y) = 0\,,
\end{equation}
where $\lambda_0$ is a constant, $x\ge 1$ and  $|y| \le 1$.  We assume henceforth that $\lambda_0 = \nu (\nu +1)$; then, $A(x)$ and $B(y)$ are both solutions of the Legendre differential equation 
\begin{equation}\label{eq: 11}
\frac{d}{dz}\left[(1-z^2)\,\frac{dF_\nu}{dz}\right] + \nu (\nu +1)\,F_\nu(z) = 0\,,
\end{equation}
where $F_{\nu}(z)$ has regular singularities at $z = \pm 1, \infty$. Let us first consider the interval $[-1, 1]$; for $z \in [-1, 1]$, the Legendre differential equation has two real linearly independent solutions: $P_\nu(z)$  is the Legendre function of the first kind and is regular for finite $z$, while $Q_\nu(z)$ is a Legendre function of the second kind and is singular at $z = \pm1$. If $\nu$ is an integer, $\nu = n \ge 0$, $P_n(z)$ reduces to a Legendre polynomial.  The only solutions of the Legendre differential equation with continuous first derivatives in $[-1, 1]$ are Legendre polynomials. However, if we allow $F_n$ to diverge at the end points of this interval, the solutions are Legendre functions of the second kind $Q_n(z)$, see Appendix A. 

Let us now return to our solution of Laplace's equation via $A(x)$ and $B(y)$, where $x\ge 1$, while 
$-1\le y \le1$. We choose for $B(y)$ the Legendre polynomial $P_n(y)$; however, for $A(x)$, $x \in [1, \infty)$, we must consider the solutions of the Legendre differential equation in the interval $[1, \infty)$. It turns out that, as before,  two real independent solutions exist: $\mathcal{P}_n(x)$ and $\mathcal{Q}_n(x)$. It is clear that we can choose $\mathcal{P}_n(x)$ to be the same polynomial as $P_n$ with  $x \in [1, \infty)$. On the other hand, we are particularly interested in the Legendre functions of the second kind in this interval and these are given by $\mathcal{Q}_n(x)$, where $\mathcal{Q}_n(x) \to 0$ as $x \to \infty$. 
These Legendre functions of the second kind are discussed in Appendix A. The solution of Laplace's equation that we adopt is therefore
\begin{equation}\label{eq: 12}
\psi (x, y) = \sum_{n=0}^\infty  (-1)^{n+1} q_n\, \mathcal{Q}_n(x) P_n(y)\,.
\end{equation}
It is now possible to determine the explicit expression for $\gamma(x, y)$ such that the resulting axisymmetric spacetime is  asymptotically flat and has a regular axis~\cite{ER, Quevedo:1989rfm}. 

With $\sigma= Gm/c^2 \ne 0$,  where $m$ is a mass parameter, $x=-1+r/m$  and $y=\cos\theta$, Eq.~\eqref{eq: 5} can be written as
\begin{equation}\label{eq: 13}
ds^2=-e^{2 \psi} \,dt^2 + e^{2\gamma-2\psi}\,\mathbb{B}\,\left(\frac{dr^2}{\mathbb{A}} + r^2\,d\theta^2\right) +  e^{-2\psi}\,\mathbb{A}\,r^2\,\sin^2\theta\,d\phi^2\,, 
\end{equation}  
where $\mathbb{A}$ and $\mathbb{B}$ are given by
\begin{equation}\label{eq: 14}
\mathbb{A} = 1 -\frac{2m}{r}\,, \qquad \mathbb{B} = 1 -\frac{2m}{r} + \frac{m^2}{r^2}\,\sin^2\theta\,.
\end{equation}
If in Eq.~\eqref{eq: 12} we let  $q_0=1$ and $q_n=0$ for $n > 0$, we find
\begin{equation}\label{eq: 15}
\psi = \frac{1}{2}\, \ln\left(\frac{x-1}{x+1}\right) = \frac{1}{2}\,\ln\mathbb{A}\,, \qquad \gamma =  \frac{1}{2}\, \ln\left(\frac{x^2-1}{x^2-y^2} \right) = \frac{1}{2}\,\ln\left(\frac{\mathbb{A}}{\mathbb{B}}\right)\,,
\end{equation}
so that Eq.~\eqref{eq: 13} reduces to the standard Schwarzschild metric with mass $m$ in spherical polar coordinates $(r, \theta, \phi)$. The quantities $q_n$ are proportional to the Newtonian multipole moments; in fact, Eq.~\eqref{eq: 12} corresponds to the expansion of the exterior Newtonian potential in terms of the multipole moments of the gravitational source.  For instance, if the constants $q_n$ vanish except for $q_0=1$ and  $q_2$, one recovers the Erez-Rosen solution~\cite{ER} for a gravitational source with mass $m$ and classical quadrupole moment proportional to $q_2$.  
                          
The Schwarzschild metric uniquely describes the  exterior vacuum field of a spherically symmetric distribution of matter. It follows that the static Schwarzschild solution endowed with an infinite set of classical moments due to axisymmetric deformations of a nonrotating spherical source would be interesting for astrophysical applications, except that astronomical sources generally rotate. As described in detail in Refs.~\cite{SKMH, HD}, powerful methods have been developed for generating solutions of Einstein's field equations of general relativity. Specifically, the HKX transformations (due to Kinnersley, Hoenselaers and Xanthopoulos) could be employed in the case under consideration here to generate \emph{stationary} axisymmetric vacuum solutions~\cite{QM1}.  In this way, one can find a generalization of the Kerr solution that contains a full set of gravitoelectric (due to mass) and gravitomagnetic (due to spin) multipole moments~\cite{Quevedo:1986nn, Q1}. That is, as in Eq.~\eqref{eq: 12}, one can first work out the exterior solution for a Schwarzschild source with a complete set of gravitoelectric multipole moments by generalizing the Erez-Rosen approach~\cite{Quevedo:1989rfm}; next, one can further generalize the result by ``rotating" it via HKX transformations to obtain a significant generalization of Kerr spacetime. Moreover,  it is also possible to  extend the resulting solution to include electric charge and the Taub-NUT parameter~\cite{QM2, Q2, Q3, Q4}. These stationary axisymmetric solutions are known as the \emph{generalized Kerr spacetimes}~\cite{Quevedo:1991zz, Bini:2009cg, Boshkayev:2012ej}.  

It is important to recognize that this class of generalized Kerr spacetimes may not be unique, since nothing similar to the black hole uniqueness theorems is known in this case.  For instance, another class of such solutions was investigated by Manko and Novikov~\cite{MaNo, Gair:2007kr, WCJ}. The interconnection between these two classes of solutions is unknown. In fact, there may be an infinite number of such distinct stationary axisymmetric solutions representing \emph{generalized black hole solutions}, that is, different possible final states of complete gravitational collapse with the same \emph{external} classical gravitoelectric and gravitomagnetic multipole moments. This circumstance would seem to be in contrast to Newtonian gravity, where the exterior gravitational potential is uniquely given by the multipole moments of the source. However, even in Newtonian gravity, the nature of the source cannot be uniquely determined by the exterior multipole moments. In connection with this degeneracy, let us mention that the external Newtonian gravitational potential  of a compact spherical distribution of mass $M$ is simply $-GM/r$, regardless of the nature of the radial density distribution inside the source. In fact, the source could be replaced by a point mass. The same problem exists in general relativity, since the exterior gravitational field of a finite spherically symmetric distribution of matter is uniquely described by the static Schwarzschild metric. Thus an infinite degree of degeneracy exists regarding the nature of the interior of the gravitational source. 

The generalized black hole solutions of general relativity theory may possibly be involved in generating the gravitational wave signals that have been recently detected~\cite{Abbott:2016blz, Abbott:2016nmj, TheLIGOScientific:2016pea, Abbott:2017gyy, Abbott:2017vtc, Abbott:2017oio, NEW}. On the other hand, perturbations of  generalized Kerr spacetimes have not been investigated. In particular, it is not known whether such perturbed systems undergo a ringdown at late times resulting in quasinormal mode oscillations of the kind that have apparently been received in recent observations~\cite{Abbott:2016blz, Abbott:2016nmj, TheLIGOScientific:2016pea, Abbott:2017gyy, Abbott:2017vtc, Abbott:2017oio, NEW}. This important issue will be addressed in the next section.

\section{Systems with Mass and Quadrupole Moment}

Static axisymmetric systems containing gravitoelectric multipole moments have been the subject of numerous investigations. A recent review of two-parameter systems with mass and quadrupole moment is contained in Ref.~\cite{FQS}. The simplest such solution appears to be the $\delta$-solution that is obtained from the Schwarzschild solution via a $\delta$-transformation. We will limit our considerations in this section to two such exact spacetimes, namely, those associated with the $\delta$-solution~\cite{Zip, Voo, Papadopoulos:1981wr, Boshkayev:2015jaa} and the Erez-Rosen solution~\cite{ER}.

\subsection{$\delta$-Metric}

Consider the $\delta$-metric given by
\begin{equation}\label{S1}
ds^2 = -\mathbb{A}^\delta \,dt^2 + \mathbb{A}^{-\delta}\, \left(\frac{\mathbb{A}}{\mathbb{B}}\right)^{\delta^2-1}\,dr^2 +  \mathbb{A}^{1-\delta}\,\left(\frac{\mathbb{A}}{\mathbb{B}}\right)^{\delta^2-1}\,r^2\,d\theta^2 +\mathbb{A}^{1-\delta}\,r^2\,\sin^2\theta\,d\phi^2\,,
\end{equation}
which should agree for 
\begin{equation}\label{S2}
\delta = 1 + q\,
\end{equation}
with the $q$-metric given in Ref.~\cite{Boshkayev:2015jaa}. Here,  $\mathbb{A}$ and $\mathbb{B}$ are defined in Eq.~\eqref{eq: 14}.  For $\delta = 1$, we recover the Schwarzschild metric. We assume throughout that $\delta > 0$ or $q > -1$. 
A general discussion of the scalar polynomial curvature invariants of the $\delta$-metric is given in Appendix B and the nature of the corresponding Weyl tensor is studied in Appendix C. We have verified from our more general results that the singularities of the $\delta$-metric are just the ones described in Ref.~\cite{Boshkayev:2015jaa}. All the singularities occur for $r \le 2m$; for $r > 2m$, the exterior field of the $\delta$-metric is singularity-free and static. Approaching the source from the exterior region $r > 2m$, we encounter a naked singularity at $r = 2m$, which can be  a null or a  timelike hypersurface. Other singularities exist interior to this hypersurface. The $\delta$-metric describes the exterior field of an oblate configuration for $q > 0$ and a prolate configuration for $-1 < q < 0$.  

In connection with volume element in the $\delta$-metric, we find
\begin{equation} \label{S3}
\sqrt{-g} = \frac{\mathbb{A}^{\delta^2-\delta}}{\mathbb{B}^{\delta^2-1}}\,r^2\,\sin \theta\,,
\end{equation}
so that for $\theta \ne 0, \pi$ and $\delta > 1$, the volume element corresponding to $r = 2m$ vanishes.  Thus in the case of an oblate configuration, the naked singularity at $r = 2m$ is essentially the ``origin" for exterior observers. 

The timelike Killing vector $\partial_t$ is such that $\partial_t \cdot \partial_t$ is $-\mathbb{A}^\delta$, which vanishes at $r=2m$ for $\delta > 0$. Thus in general $r = 2m$ is the static limit surface and, as expected, an infinite redshift surface, which we theoretically demonstrate towards the end of this subsection.   Next, consider the $r = $~constant hypersurface and its normal vector $N$, 
\begin{equation}\label{S4}
N\cdot N = g^{rr} = \mathbb{A}^{-\delta^2 + \delta +1}\,\mathbb{B}^{\delta^2-1}\,.
\end{equation}
For $\theta = 0, \pi$, the $r = 2m$ hypersurface is always null. Moreover,  for $\theta \ne 0, \pi$, the hypersurface is again null for $\delta +1 > \delta^2$ or $1-q > q^2$ for 
$-(\sqrt{5} + 1)/2 < q < (\sqrt{5}-1)/2$.
It follows that the hypersurface $r = 2m$ is always null for a prolate configuration. For an oblate configuration, the $r = 2m$ hypersurface is null provided  $q < (\sqrt{5}-1)/2$; otherwise, the $r = 2m$ hypersurface is timelike (except at $\theta = 0, \pi$, where it is again null). This means that for all prolate configurations as well as for  oblate configurations with $q \lesssim 0.6$, the null singular hypersurface $r = 2m$ acts as a one-way membrane, so that one can set up ingoing boundary conditions for QNMs in this case just as in the case of a black hole. It is therefore possible that with this limitation on the magnitude of $q$ for oblate systems, the notion of quasinormal oscillations makes sense for such configurations.

\subsubsection{Physical Interpretation of the $\delta$-Metric}

Let us start with the $\delta$-metric given by Eq.~\eqref{S1} and let $\delta = 1 + q$, where we recover the Schwarzschild metric for $q=0$; therefore, we may think of $q$ as a deformation parameter for the Schwarzschild black hole. For small deformations, we may expand the metric to linear order in $q$. The result can be expressed as
\begin{equation}\label{P1}
-g_{tt} = \left(1-\frac{2m}{r}\right)\, \left[1+ q\,\ln\left(1-\frac{2m}{r}\right)\right]\,,
\end{equation}
\begin{equation}\label{P2}
g_{rr} = \frac{1}{1-\frac{2m}{r}}\, \left[1- q\,\ln\left(1-\frac{2m}{r}\right) - 2 q\,\ln\left(1+ \frac{m^2 \sin^2\theta}{r^2-2mr}\right)\right]\,,
\end{equation}
\begin{equation}\label{P3}
g_{\theta \theta} = r^2\, \left[1- q\,\ln\left(1-\frac{2m}{r}\right) - 2 q\,\ln\left(1+ \frac{m^2 \sin^2\theta}{r^2-2mr}\right)\right]\,,
\end{equation}
\begin{equation}\label{P4}
g_{\phi \phi} = r^2 \sin^2\theta\,\left[1-  q\, \ln\left(1-\frac{2m}{r}\right)\right]\,,
\end{equation}
where we have used the fact that for $Z > 0$,
\begin{equation}\label{P5}
Z^q = e^{q \ln{Z}} = 1 + q \ln{Z} + O(q^2)\,.
\end{equation}

Let us consider next the coordinate transformations
\begin{equation}\label{P6}
r = \rho\,\left[1- q\,\frac{m}{\rho} - q\,\frac{m^2}{\rho^2}\left(1 + \frac{m}{\rho} + \cdots \right)\,\sin^2\vartheta \right]\,
\end{equation}
and
\begin{equation}\label{P7}
\theta = \vartheta - q\,\frac{m^2}{\rho^2}\left(1 + 2 \frac{m}{\rho} + \cdots \right)\,\sin\vartheta \cos\vartheta\,,
\end{equation}
where we have neglected terms of order $m^4/\rho^4$. Implementing these transformations in the metric coefficients and expanding terms such as $\ln(1-2m/\rho)$ to order $m^3/\rho^3$, namely, 
\begin{equation}\label{P8}
\ln\left(1-\frac{2m}{\rho}\right) = 1- 2\,\frac{m}{\rho} - 2\, \frac{m^2}{\rho^2} - \frac{8}{3}\,\frac{m^3}{\rho^3} -\cdots\,,
\end{equation}
we find that the spacetime metric takes the form
\begin{equation}\label{P9}
ds^2 = - c^2\,(1+ 2\, \Phi_N/c^2) \,dt^2 + \frac{d\rho^2}{1+ 2\,\Phi_N/c^2} + U(\rho, \vartheta) \rho^2 (d\vartheta^2 + \sin^2\vartheta\,d\phi^2)\,,
\end{equation}
where $\Phi_N$ is the Newtonian potential involving mass $M$ and quadrupole moment $Q$. More precisely,
\begin{equation}\label{P10}
 \Phi_N = -\frac{GM}{\rho}  + \frac{GQ}{\rho^3}\,P_2(\cos\vartheta)\,, \qquad U(\rho, \vartheta) = 1 - 2\,\frac{GQ}{c^2\rho^3}\,P_2(\cos\vartheta)\,,
\end{equation}
where $P_2(\cos\theta)= (3\cos^2\theta-1)/2$ is a Legendre polynomial with $P_2(\pm1) = 1$ along the symmetry axis and $P_2(0) = -1/2$ in the equatorial plane . Moreover, we have 
\begin{equation}\label{P11}
 M = (1+ q) \,m\,, \qquad Q = \frac{2}{3} m^3\,q\,. 
\end{equation}
These physical parameters of the system under consideration in this work agree with invariant definitions of multipole moments in general relativity to first order in $q$~\cite{Boshkayev:2015jaa, FAS}.

We note that for an oblate system $Q > 0$. Writing $Q = M \bar{r}^2$, where $\bar{r}$ is a lengthscale  characteristic of the system under consideration, the positive dimensionless parameter $q$ can be expressed as
\begin{equation}\label{P12}
 q = 6 \left(\frac{2GM}{c^2\, \bar{r}}\right)^{-2}\,. 
\end{equation}
We treat the quadrupole moment to first order in this paper; therefore, $\bar{r} = \sqrt{Q/M}$  should be such that $\bar{r} \ll 2GM/c^2$.

The Einstein tensor for metric~\eqref{P9} has dominant terms of order $q\,m^4/\rho^4$, which are the main terms that we neglected in our analysis. Thus despite its deceptive form, metric~\eqref{P9} is simply a weak-field post-Newtonian limit of the $\delta$-metric.

\subsubsection{Quadrupole Moment}

Regarding the definition of the quadrupole moment, let us recall that in Newtonian gravity we have for the external potential of a matter distribution
\begin{equation}\label{Q1}
 \Phi_N (\mathbf{x}) = - G \int \frac{\rho_{matter}(x')\,d^3x'}{|\mathbf{x} - \mathbf{x'}|}\,,
\end{equation}
where $\rho_{matter}$ is the density of the distribution of matter in a compact body and $\mathbf{x}$ is the position vector of a point exterior to the body starting from an origin inside the body. Let $r = |\mathbf{x}|$ and 
$r' = |\mathbf{x'}|$, then for $r > r'$, we can write
\begin{equation}\label{Q2}
 \Phi_N (\mathbf{x}) = - G\,\frac{M}{r} - G\,\frac{\mathbf{D}\cdot \mathbf{x}}{r^3} - G\, \frac{Q_{ij} \,x^ix^j}{r^5} - \cdots \,,
\end{equation}
where the dipole moment and the symmetric and traceless quadrupole tensor of the system are given by 
\begin{equation}\label{Q3}
\mathbf{D} =  \int \rho_{matter}(x')\,\mathbf{x'}\,d^3x'\,, \qquad Q_{ij} =  \int \rho_{matter}(x')\,(3 x'_i x'_j - x'^2 \delta_{ij})\,d^3x'\,.
\end{equation}
In general, $\mathbf{D}$ vanishes if the origin of coordinates coincides with the Newtonian center of mass of the source. On the other hand, if the system possesses axial symmetry about the chosen origin and is also reflection symmetric about its equatorial plane, then $\mathbf{D} = 0$. For such a system the quadrupole tensor is then diagonal in Cartesian coordinates and is given by ${\rm diag}(Q_{11}, Q_{22}, Q_{33})$, where $Q_{33} = -2\,Q_{11} = -2\,Q_{22}$. Let us define the quadrupole moment of such a system via
\begin{equation}\label{Q4}
Q := - Q_{33} = 2\,Q_{11} = 2\,Q_{22}\,.
\end{equation}
It is then possible to write  
\begin{equation}\label{Q5}
 \Phi_N (\mathbf{x}) = - G\,\frac{M}{r} + G\, \frac{Q}{r^3}\,\frac{3\cos^2\theta-1}{2} - \cdots \,,
\end{equation}
where the Cartesian coordinates $(x, y, z)$ are expressed as $(r, \theta, \phi)$ in spherical polar coordinates. 

If in the definition of the quadrupole tensor we replace the matter density by a certain \emph{constant} average density, then it is straightforward to show that $Q := - Q_{33}$ is \emph{positive} for an \emph{oblate} system. It is negative for a prolate system and vanishes for a spherical system. Thus for fixed $r$, the quadrupolar contribution to the Newtonian potential is negative in the direction of elongation and positive in the opposite direction.

\subsubsection{Gravitational Shift of Frequency in $\delta$-Spacetime}

It is interesting to consider the gravitational frequency shift in $\delta$-spacetime. For the sake of simplicity, we consider \emph{radial} null geodesics such that along the world line we have $\theta$ = constant and $\phi$ = constant. The geodesic equations of motion can be simply obtained from a Lagrangian of the form $(ds/d\lambda)^2$, where $\lambda$ is an affine parameter along the world line. Using this approach, we find that in the case of the $\delta$-metric, we have $dt/d\lambda = C_0/(- g_{tt})$, 
where $C_0 > 0$ is a constant. This result can also be obtained from the projection of the 4-velocity of the null geodesic $k^\mu = dx^\mu/d\lambda$ on the timelike Killing vector of the $\delta$-metric. From $k \cdot k = 0$, we find $dr / d\lambda = \pm C_0 / (- g_{tt}\,g_{rr})^{1/2}$, where a plus (minus) sign indicates an outgoing (ingoing) null ray. 

Imagine a static observer that occupies a fixed spatial position along the path of the null geodesic ray in the exterior $\delta$-spacetime. The observer's 4-velocity is given by $u = (- g_{tt})^{-1/2}\partial_t$ and the frequency of light propagating along the radial direction as measured by the observer is $\omega_r = -u \cdot k = \omega_{\infty}\,(-g_{tt})^{-1/2}$, 
where $\omega_{\infty} = C_0$ is the frequency of light as measured by static inertial observers at spatial infinity along the ray. In this way, one can determine the contribution of the quadrupole moment to the gravitational shift of the frequency of light. More specifically,  $\omega_r = C_0\,\mathbb{A}^{-\delta/2}$,
so that as the static observer approaches the naked singularity, $\mathbb{A} \to 0$,  and the measured frequency diverges. That is, the $r= 2m$ singularity is an infinite redshift surface for outgoing null rays.  

For further recent studies of the $\delta$-metric, see~\cite{Quevedo:2016gfl, Quevedo:2016rsw, Abishev:2015pma, Luongo:2014qoa, Toktarbay:2015lua, Quevedo:2010mn} and the references cited therein. We note that the Tomimatsu-Sato metric reduces in the absence of rotation to the $\delta$-metric for integer $\delta$~\cite{SKMH,Griffiths:2009dfa,Papadopoulos:1981wr}. 
Moreover, a rotating $\delta$-spacetime is contained in Ref.~\cite{Toktarbay:2015lua}. 

\subsection{Erez-Rosen Metric}

The work of Erez and Rosen~\cite{ER} extended the method of Weyl and Levi-Civita to find the exterior gravitational field of a Schwarzschild source possessing a multipole moment. In particular, they explicitly worked out the case of an object with mass parameter $m$ and quadrupole parameter $q_2$. The Erez-Rosen solution~\eqref{eq: 5} is given in prolate spheroidal coordinates by
\begin{equation}\label{K1}
\psi  = \frac{1}{2} \ln\left(\frac{x-1}{x+1}\right)+ q_2\,\frac{1}{2} (3y^2-1)\left[\frac{3x^2-1}{4}\,\ln\left(\frac{x-1}{x+1}\right) + \frac{3}{2}\,x \right]\,,
\end{equation}
\begin{align}\label{K2}
\gamma  = {} & \frac{1}{2}(1+q_2)^2 \ln\left(\frac{x^2-1}{x^2-y^2} \right) -\frac{3}{2}\,q_2 (1-y^2)\,\left[ x\, \ln\left(\frac{x-1}{x+1}\right) + 2 \right] \\
\nonumber & + \frac{9}{16}\,q_2^2 (1-y^2)\left[x^2 +4y^2 -9x^2y^2-\frac{4}{3} +\left(x^2 + 7 y^2 -9x^2y^2 
-\frac{5}{3}\right)\,x\, \ln\left(\frac{x-1}{x+1}\right)\right] \\
\nonumber & + \frac{9}{64}\,q_2^2 (x^2-1) (1-y^2)(x^2+y^2-9x^2y^2-1)\,\ln^2\left(\frac{x-1}{x+1} \right)\,.
\end{align}
With $x+1 = r/m$ and $y = \cos \theta$, we obtain the Schwarzschild solution augmented by a quadrupole moment. In fact, for $m \ll r$ and to first order in $q_2$, one can write in the weak-field approximation
\begin{equation}\label{K3}
-g_{tt} = e^{2\psi}  =  \left(1 -\frac{2m}{r}\right)\,\left\{1 - \frac{4}{15}\,q_2\,\frac{m^3}{r^3}\,P_2(\cos\theta)\,\left[1 + O\left(\frac{m}{r}\right)\right]\right\}\,,
\end{equation}
which may be compared to $-g_{tt} = 1+ 2\,\Phi_N$ in the Newtonian limit of general relativity. Therefore, for the Erez-Rosen solution we have
\begin{equation}\label{K4}
M = m\,, \qquad Q = -\frac{2}{15}\,q_2\,m^3\,.
\end{equation}
Thus $q_2 < 0$ for an oblate configuration and $q_2 > 0$ for a prolate configuration. 

Approaching the $r= 2m$ hypersurface from the exterior region, the timelike Killing vector becomes null at $r = 2m$ provided $1+q_2\,P_2(\cos\theta) > 0$, in which case the $r=2m$ is a static limit surface. This is always the case if $|q_2|$ is sufficiently small compared to unity.  A discussion of the gravitational frequency shift along radial null geodesics in the Erez-Rosen spacetime is contained in Ref.~\cite{MaQu}.

As expected, the $r = 2m$ hypersurface is a naked singularity~\cite{Quevedo:1991zz}. Let $N =g^{rr} \partial_r$ be the vector normal to the $r$ = constant$~ > 2m$ hypersurface. Then, $N \cdot N = g^{rr}$. For $\theta = 0, \pi$, we find that as $r \to 2m$, $x \to 1$, the $r= 2m$ hypersurface is null for $1+q_2>0$, but it is otherwise timelike. This means that the hypersurface is null for $Q/M^3 < 2/15$. On the other hand, for $\theta \ne 0, \pi$, we find that as $r \to 2m$ and $x \to 1$, 
\begin{equation}\label{K5}
g^{rr} \sim (x-1)^{\varpi}\,,
\end{equation}
where
\begin{equation}\label{K6}
\varpi = 1 +\bar{\delta} - \bar{\delta}^2\,, \qquad  \bar{\delta} = \frac{15Q}{2M^3}\,P_2(\cos\theta)\,.
\end{equation}
Let us first suppose that $P_2(\cos\theta) = 1-\frac{3}{2}\,\sin^2\theta = 0$. This occurs for $\theta_c$ and $\pi - \theta_c$, where  $\theta_c\approx 54.7^\circ$ is such that $\sin\theta_c = \sqrt{2/3}$.  For $\theta = \theta_c$ and $\theta = \pi - \theta_c$, we find from Eq.~\eqref{K6} that $\varpi = 1$ and the $r= 2m$ hypersurface is null. On the other hand, suppose that $P_2(\cos\theta) \ne 0$; then, it follows from Eq.~\eqref{K6} that  $\varpi > 0$ and the $r=2m$ hypersurface is therefore again null for
\begin{equation}\label{K7}
-\frac{\sqrt{5}-1}{2} < \frac{15Q}{2M^3}\,P_2(\cos\theta) < \frac{\sqrt{5}+1}{2}\,.
\end{equation}
Otherwise, the $r=2m$ hypersurface is timelike. For instance, in the equatorial plane, the $r = 2m$ hypersurface is null provided
\begin{equation}\label{K8}
-\sqrt{5} -1 < \frac{15Q}{2M^3} < \sqrt{5} -1\,.
\end{equation}
More generally, we note that for $0 < \theta < \theta_c$ and $\pi - \theta_c < \theta < \pi$, $0 < P_2(\cos\theta) < 1$, above and below the equatorial plane, while  for $\theta_c < \theta < \pi - \theta_c$ around the equatorial plane, $-\frac{1}{2} < P_2(\cos\theta) < 0$.

A remark is in order here regarding an essential similarity of this Erez-Rosen case with that of the $\delta$-metric. The naked singularity at $r= 2m$ is null and acts as a one-way membrane if $1+\bar{\delta} > \bar{\delta}^2$, where $\bar{\delta} :=  -q_2 \,P_2(\cos\theta)$, while in the case of the $\delta$-metric, we have $1+\delta > \delta^2$, where $\delta = 1+q$. In the latter case, we expect QNMs in the case of oblate systems with $0 < q \lesssim 0.6$.  In our approach to QNMs of such systems, the quadrupole moment is treated to linear order of approximation; therefore, for oblate configurations in the Erez-Rosen case, $q_2 < 0$ and when the quadrupole moment is sufficiently small  Eq.~\eqref{K7} is always satisfied for $0 < \theta < \pi$.

\section{Perturbative Treatment of the Quadrupole Moment}

In the exact $(M, Q)$ solutions discussed thus far, the spacetime singularities occur for $r \le 2m$; indeed, approaching the collapsed system from the exterior region one first encounters a naked singularity at $r = 2m$~\cite{Quevedo:1991zz}. We have shown that this hypersurface is null and acts as a one-way membrane for sufficiently small $|Q|/m^3$; henceforth, we treat the quadrupole moment to linear order of approximation and expect QNM oscillations to occur when such post-Schwarzschild systems undergo massless scalar, electromagnetic or gravitational perturbations. 

In the present section, we consider the post-Schwarzschild approximation involving quadrupole moment to linear order represented by two spacetimes: the Schwarzschild-Quadrupole ($SQ$) spacetime, which is  a variant of the $\delta$-spacetime, and the static Hartle-Thorne ($HT$) spacetime~\cite{HaTh}, which, as we show in the last part of this section, is related to the Erez-Rosen spacetime.  In fact, starting from the Erez-Rosen solution, we obtain to linear order in the quadrupole moment the static Hartle-Thorne solution via a certain $\delta$-transformation.

\subsection{$SQ$-Metric}

We are interested in the strong-field regime near a Schwarzschild black hole that has a quadrupole moment. The latter is taken into account to linear order in our treatment for the sake of simplicity. The simplest known solution of GR suitable for our purpose is the 
$\delta$-metric given by Eq.~\eqref{S1} that has two parameters $m$ and $\delta$. In view of Eq.~\eqref{P11}, we can replace $\delta$ by $1+q$,  $m$ by $M/(1+q)$ and then linearize the resulting metric in the dimensionless quadrupole parameter $q$. To first order in $q$, the quadrupole moment is then given by $Q$, where
\begin{equation}\label{R1}
 Q = \frac{2}{3} M^3\,q\,.
\end{equation}
The end result is $ds^2_{SQ} = g_{\mu \nu}\, dx^\mu \,dx^\nu$, 
where $g_{\mu \nu}$ is diagonal and represents the Schwarzschild spacetime of mass $M$ perturbed by terms that are linear in the quadrupole moment $Q$. As expected, the Einstein tensor for this metric is proportional to $q^2$ and vanishes when terms proportional to $q^2$ and higher are neglected. It is useful to write the $SQ$-metric in the following form
\begin{align}\label{R6}
ds_{SQ}^2 ={}& -  \left[1+q\left(\frac{2M}{r\mathbb{A}} +\ln{\mathbb{A}}\right)\right]\,\mathbb{A}\,dt^2+  \left[1-q\left(\frac{2M}{r\mathbb{A}} +\ln{\frac{\mathbb{B}^2}{\mathbb{A}}}\right)\right]\,\frac{dr^2}{\mathbb{A}}\\
\nonumber & + \left(1-q\ln{\frac{\mathbb{B}^2}{\mathbb{A}}}\right)\,r^2\,d\theta^2 + (1-q\ln{\mathbb{A}})\,r^2\sin^2\theta\,d\phi^2\,.
\end{align}
The connection coefficients for the $SQ$-metric can be obtained from the results given in Appendix D.

\subsection{Static Hartle-Thorne Metric}

The rotating Hartle-Thorne solution~\cite{HaTh} represents the exterior vacuum spacetime domain that matches smoothly to a realistic model of a slowly rotating relativistic star of mass $M$, angular momentum $J$ and quadrupole moment $Q$. The latter two parameters are treated to second and first orders of approximation, respectively. In this section we set $J=0$ and consider the static Hartle-Thorne ($HT$) spacetime. Though formally more complicated than the $SQ$-metric, the static $HT$-metric appears to be more suitable for physical applications.  

The static Hartle-Thorne metric for a mass $M$ with quadrupole moment $Q$, the latter treated to linear order, is given by~\cite{HaTh}
\begin{equation}\label{H1}
ds_{HT}^2 = -\mathcal{F} \,dt^2 + \frac{1}{\mathcal{F}}\,dr^2 +  \mathcal{G}\,r^2\,(d\theta^2 +\sin^2\theta\,d\phi^2)\,,
\end{equation}
where
\begin{equation}\label{H2}
\mathcal{F} = \left(1 -\frac{2M}{r}\right)\,\left[ 1 + \frac{5Q}{4M^3}\, \mathcal{Q}_2^2\left(\frac{r}{M}-1\right)\,P_2(\cos\theta)\right]\,
\end{equation}
and
\begin{equation}\label{H3}
\mathcal{G} = 1 +  \frac{5Q}{4M^3}\,\left[\frac{2M}{\sqrt{r(r-2M)}}\,\mathcal{Q}_2^1\left(\frac{r}{M}-1\right)-\mathcal{Q}_2^2\left(\frac{r}{M}-1\right)\right]\,P_2(\cos\theta)\,.
\end{equation}
Here, $\mathcal{Q}_n^m(x)$ is the associated Legendre function of the second kind for $x \in [1, \infty)$, namely, 
\begin{equation}\label{H4}
\mathcal{Q}_n^m(x) =  (-1)^m\,(x^2-1)^{m/2}\,\frac{d^m\mathcal{Q}_n(x)}{dx^m}\,. 
\end{equation}
From the formula for $\mathcal{Q}_2(x)$ given in Appendix A, we find
\begin{equation}\label{H5}
\mathcal{Q}_2^1(x) =  - (x^2-1)^{1/2}\,\left[\frac{3}{2} \,x\,  \ln\left(\frac{x+1}{x-1}\right) -\frac{3x^2-2}{x^2-1}\right]\,,
\end{equation}
\begin{equation}\label{H6}
\mathcal{Q}_2^2(x) =  \frac{3}{2}\, (x^2-1)\,\ln\left(\frac{x+1}{x-1}\right) -x\,\frac{3x^2-5}{x^2-1}\,,
\end{equation}
etc. 

For $M\ll r$,
\begin{equation}\label{H7}
\mathcal{Q}_2^1\left(\frac{r}{M}-1\right) =  \frac{2\,M^3}{5\,r^3}\left[1 + O\left(\frac{M}{r}\right)\right]\,,
\end{equation}
\begin{equation}\label{H8}
\mathcal{Q}_2^2\left(\frac{r}{M}-1\right) =  \frac{8\,M^3}{5\,r^3}\left[1 + O\left(\frac{M}{r}\right)\right]\,,
\end{equation}
etc. Therefore, 
\begin{equation}\label{H9}
\frac{2M}{\sqrt{r(r-2M)}}\,\mathcal{Q}_2^1\left(\frac{r}{M}-1\right)-\mathcal{Q}_2^2\left(\frac{r}{M}-1\right) =  - \frac{8\,M^3}{5\,r^3}\left[1 + O\left(\frac{M}{r}\right)\right]\,.
\end{equation}
 If we now assume that $M \ll r$  in the Hartle-Thorne metric and expand the relevant radial quantities in powers of $M/r$, we find using Eqs.~\eqref{H7}--\eqref{H9} that the result coincides with the post-Newtonian metric given in  Eqs.~\eqref{P9} and~\eqref{P10}. This circumstance justifies the designation of $M$ and $Q$ as the mass and the quadrupole moment of the gravitational source in this case. 

The Einstein tensor for the static Hartle-Thorne solution turns out to be proportional to $Q^2$ and vanishes when such terms are neglected. In this solution the mass is thus taken into account to all orders, while the quadrupole moment is only considered to linear order.

\subsubsection{Relation Between Erez-Rosen and Static Hartle-Thorne Solutions}

It is possible to establish a connection between the Erez-Rosen~\cite{ER} and Hartle-Thorne~\cite{HaTh} solutions in the absence of angular momentum, see in this connection Appendix B of Ref.~\cite{MaTh}. To this end, one can first employ the $\delta$-transformation to obtain a new class of solutions from the Erez-Rosen solution. In this class of solutions, we let $\delta = 1 + \sigma_0\,q_2$, where $\sigma_0$ is a real parameter and consider the class of solutions to linear order in $q_2$ that contain parameters $(M, q_2, \sigma_0)$. Let us next consider the transformations given by $M = M' (1+q_2)$ and $(r, \theta) \mapsto (r', \theta')$, where  
\begin{equation}\label{H10}
r = r' + M' \,q_2 +\frac{3}{2}\,M' \,q_2\, \left[x' +\frac{1}{2}(x'^2-1)\, \ln\left(\frac{x'-1}{x'+1}\right)\right]\,\sin^2\theta'\,
\end{equation}
and
\begin{equation}\label{H11}
\theta = \theta' - \frac{3}{2} \,q_2\, \left[2+ x'\, \ln\left(\frac{x'-1}{x'+1}\right)\right]\,\sin\theta'\,\cos\theta'\,.
\end{equation}
Here, $x' = -1 + r' / M'$. To first order in $q_2$, we recover the metric of the static Hartle-Thorne solution~\cite{HaTh} for 
\begin{equation}\label{H12}
\sigma_0 = -1\,, \qquad M_{HT} = M'\,, \qquad Q_{HT} = -\frac{4}{5}\, M'^3\, q_2\,.
\end{equation}
It follows that the connections between the Erez-Rosen and static Hartle-Thorne parameters are given by
\begin{equation}\label{H13}
M_{HT} = M_{ER} (1-q_2)\,, \qquad Q_{HT} = 6\,Q_{ER}\,.
\end{equation}

The above connection between the two metrics can be further generalized by taking angular momentum into account to second order~\cite{Quevedo:2010vx}.  More specifically, Quevedo~\cite{Quevedo:2010vx} has extended the above transformations to second order in angular momentum and has thereby  established a connection between the Quevedo-Mashhoon metric~\cite{QM1} and the Hartle-Thorne metric~\cite{HaTh} in the presence of angular momentum. For further work regarding the Hartle-Thorne solution, see Ref.~\cite{Frutos-Alfaro:2015lua}.

Finally, the Christoffel symbols of the static Hartle-Thorne ($HT$) solution can be obtained for $J = 0$ from those of the rotating Hartle-Thorne ($RHT$)  solution presented in Appendix D.

\section{Quasinormal Modes}

The results of Section IV will be useful for the approximate analytic calculation of the frequencies of \emph{certain} QNMs of Schwarzschild black hole with quadrupole moment in the eikonal limit via the light-ring method. Numerical approaches are needed to do a more complete investigation. 

\subsection{General Description of the Analytic Light-Ring Method}

The early analytic approaches to the quasinormal modes (QNMs) of black holes~\cite{Ma1, BlMa, FeMa1, FeMa2, Ma2, LiMa} included a method that is \emph{independent of the separability of massless (scalar, electromagnetic and gravitational) wave equations} on black hole backgrounds. This method was applied to Kerr and Kerr-Newman black holes in Refs.~\cite{FeMa1}--\cite{FeMa2} and~\cite{Ma2}, respectively.  Only certain QNMs are accessible using this approach~\cite{FeMa1, FeMa2}. The purpose of this subsection is to provide a brief explanation of this method; in subsequent subsections, we will apply this approach to find QNMs of a perturbed Schwarzschild black hole with  quadrupole moment using the $SQ$-metric as well as the $HT$-metric.  The former is the simplest to deal with analytically, while the latter has desirable  physical properties. Furthermore, we have shown that, to  linear order in the quadrupole moment, the Erez-Rosen solution can be essentially transformed to the static Hartle-Thorne solution.  

To calculate QNMs, one may employ any appropriate perturbation and deduce the corresponding QNMs from the temporal evolution of the perturbation, since the QNMs constitute intrinsic properties of black holes and are independent of any particular perturbation. Thus we consider the bundle of null rays in the unstable circular equatorial orbit  at $r = r_0$ around the Schwarzschild black hole with quadrupole moment and study the evolution of the rays as they propagate outward to spatial infinity and inward toward the central collapsed configuration. The unperturbed null circular orbit can be written as
\begin{equation}\label{M1}
\bar{x}^\mu = (t, r, \theta, \phi) = (\lambda, r_0, \frac{\pi}{2}, \omega_{\pm}\,\lambda)\,,
\end{equation}
where $\lambda$ is an affine parameter along the null geodesic orbit. The corresponding propagation vector is thus $k^\mu$, 
\begin{equation}\label{M2}
k^\mu = \frac{d\bar{x}^\mu}{d\lambda} = (1, 0, 0, \omega_{\pm})\,
\end{equation}
in the $(t, r, \theta, \phi)$ coordinate system. Here, $r_0$ is the radius of the unperturbed orbit and $\omega_{\pm}$ are frequencies associated with the orbit such that $\omega_{+}$ refers to   corotating and $\omega_{-}$ refers to counter-rotating motions. 

In the static axisymmetric background spacetimes under consideration, the massless wave perturbations are superpositions of a set of eigenmodes of the form 
\begin{equation}\label{M3}
e^{i (\omega\, t- \mu\,\phi)}\, \mathcal{S}_{\omega j \mu s}(r, \theta)\,,
\end{equation}
which takes into account the Killing symmetries of the background spacetime. The wave eigenmode~\eqref{M3} has frequency $\omega$, total angular momentum parameters $(j, \mu)$, with  $-j \le \mu \le j$, and spin $s$. The null rays correspond to the eikonal limit of wave motion such that $M\, \omega \gg 1$; moreover, the circular null rays are characterized by $|\mu| = j \gg 1$. That is, $\mu = j$ for corotating rays, while the counter-rotating rays are characterized by $\mu = -j$. In this approach, the resulting QNMs are independent of the spin of the massless wave perturbation. It follows from Eqs.~\eqref{M1} and~\eqref{M3} that for the null circular orbits, the dominant perturbation frequency according to static inertial observers at spatial infinity is given by 
\begin{equation}\label{M4}
\omega^0 = \mu\,\frac{d\phi}{dt} =  \pm j\,\omega_{\pm}\,.
\end{equation}

Next, we consider a slight perturbation of the circular equatorial null orbit  at $t = 0$. The perturbed orbit is then given by $\bar{x}^\mu \mapsto x^\mu$, where 
\begin{equation}\label{M5}
r = r_0 [ 1 + \epsilon\,f(t)]\,, \qquad \phi = \omega_{\pm} [t + \epsilon\,g(t)]\,, \qquad \lambda = t + \epsilon\,h(t)\,.
\end{equation}
Here,  $\epsilon$ is a dimensionless perturbation parameter,  $0 < |\epsilon| \ll 1$.  The perturbed orbit~\eqref{M5} should be substituted in the null geodesic equations
to determine the functions $f(t)$, $g(t)$ and $h(t)$ such that 
\begin{equation}\label{M7}
f(0) = g(0) = h(0) = 0\,.
\end{equation}

It will turn out from the null path equation, $g_{\mu \nu} dx^\mu\,dx^\nu = 0$, that $g(t) = 0$, so that 
\begin{equation}\label{M8}
\frac{d\phi}{dt} = \omega_{\pm} + O(\epsilon^2)\,.
\end{equation}
This means that to $O(\epsilon^2)$, the \emph{real parts of the corresponding QNM frequencies} are given by Eq.~\eqref{M4}. 

The divergence of the null rays away from the unperturbed orbit corresponds to the decay of the QNM wave amplitude with time in the eikonal limit.  The imaginary parts of the QNM frequencies  are therefore  given by the decay rate of the orbit. The rate of divergence of nearby trajectories can also be characterized via Lyapunov exponents; for the connection between the decay rate of QNM frequencies with the corresponding Lyapunov exponents, see Ref.~\cite{Cardoso:2008bp}. For a congruence of null rays along the perturbed null equatorial circular orbit, the law of conservation of rays is given by
\begin{equation}\label{M9}
(\rho_n\,K^\mu)_{;\mu} = 0\,,
\end{equation}
where $\rho_n$ is the density of null rays and $K^\mu$ is the perturbed propagation vector given to $O(\epsilon^2)$ by
\begin{equation}\label{M10}
K^\mu = \frac{dx^\mu}{d\lambda} = [ 1 - \epsilon h', \epsilon r_0 f', 0, \omega_{\pm}\,(1 - \epsilon h')]\,,
\end{equation}
where $f' = df/dt$, etc. As we will demonstrate in detail below, the imaginary parts of the corresponding QNM frequencies can then be deduced from
\begin{equation}\label{M11}
\frac{1}{\rho_n}\,\frac{d\rho_n}{d\lambda}= -K^{\mu}{}_{;\mu} = -\frac{1}{\sqrt{-g_{MQ}}}\,\frac{\partial}{\partial x^\alpha}(\sqrt{-g_{MQ}}\,K^\alpha)\,,
\end{equation}
where $g_{MQ}$ is the determinant of the spacetime metric of a perturbed Schwarzschild black hole representing a collapsed configuration with mass $M$ and quadrupole moment $Q$. Note that for $\epsilon = 0$, $K^\alpha$ reduces to $k^\alpha$ and hence the right-hand side of Eq.~\eqref{M11} vanishes, since the metric is independent of time $t$ and the azimuthal coordinate $\phi$; therefore, the density of null rays $\rho_n$ remains constant. However, for $\epsilon \ne 0$, $\rho_n$ decays, as will be demonstrated explicitly in the next two subsections. 

For recent studies related to this method, see Refs.~\cite{McWilliams:2018ztb, Assumpcao:2018bka, Konoplya:2017wot} and the references cited therein. For more recent general review articles on QNMs, see Refs.~\cite{Ferrari:2007dd, Berti:2009kk, Konoplya:2011qq}.

\subsection{QNMs of the $SQ$ Spacetime}

\subsubsection{Light Ring for $SQ$ Spacetime}

Let $x^{\mu}(\lambda)$ represent a null geodesic of the $SQ$-metric~\eqref{R6}, where, as before,  $\lambda$ is an affine parameter along the null world line. Then,  the equation for a null path ($ds^2 = 0$) and the geodesic equation must be satisfied; that is, 
\begin{equation}\label{R7}
g_{\mu \nu}\, \frac{dx^\mu}{d\lambda}\, \frac{dx^\nu}{d\lambda} = 0\,,\qquad \frac{d^2x^\alpha}{d\lambda^2} + \Gamma^\alpha_{\mu \nu}\, \frac{dx^\mu}{d\lambda}\, \frac{dx^\nu}{d\lambda} = 0\,.
\end{equation}

The spherically symmetric Schwarzschild spacetime has an unstable photon sphere at $r_{Sch} = 3M$. The presence of the quadrupole moment modifies this situation. That is, due to the  axial and reflection symmetries of the $SQ$-metric, we expect the existence of an unstable circular null orbit in the equatorial plane. For such an orbit with constant radial coordinate $r_0$ and $\theta = \pi/2$, $ds^2=0$ implies  
\begin{equation}\label{R8}
\left(\frac{d\phi}{dt}\right)^2 = -\frac{g_{tt}}{g_{\phi \phi}}\,.
\end{equation}
On the other hand, the geodesic equation simplifies in this case and the main contribution comes from the radial component of this equation which implies 
\begin{equation}\label{R9}
\left(\frac{d\phi}{dt}\right)^2 = -\frac{g_{tt, r}}{g_{\phi \phi, r}}\,,
\end{equation}
where we have used the fact that 
\begin{equation}\label{R10}
\Gamma^r_{tt} = \frac{1}{2} g^{rr}(-g_{tt, r})\,, \qquad \Gamma^r_{\phi \phi} = \frac{1}{2} g^{rr}(-g_{\phi \phi, r})\,.
\end{equation}
Using the other components of the geodesic equation, one can show that the orbit has the form $\bar{x}^\mu(\lambda) = (\lambda, r_0, \pi/2, \omega_{\pm}\,\lambda)$, where $d\phi/dt = \omega_{\pm}$, $\omega_{+}$ indicates the constant frequency of the corotating circular orbit and $\omega_{-}$ indicates the constant frequency of the counter-rotating orbit. 

We therefore find explicitly from Eqs.~\eqref{R8} and~\eqref{R9}, 
\begin{equation}\label{R11}
\left(\frac{d\phi}{dt}\right)^2 = \frac{\mathbb{A}}{r^2}\left [1 + 2q\left(\frac{M}{r\mathbb{A}} +\ln{\mathbb{A}}\right)\right]\,
\end{equation}
and 
\begin{equation}\label{R12}
\left(\frac{d\phi}{dt}\right)^2 = \frac{M}{r^3}\left[1 + q\left(\frac{M}{r\mathbb{A}} + 2\ln{\mathbb{A}}\right)\right]\,,
\end{equation}
respectively. These relations imply
\begin{equation}\label{R13}
r_0 = 3M \left(1 - \frac{1}{3}\,q\right)\,, \qquad \frac{d\phi}{dt} = \omega_{\pm} =  \pm \frac{1}{3\sqrt{3}\,M}\, [1 - q\,(-1+\ln{3})]\,,
\end{equation}
where  $q = 3Q/(2M^3)$.  Using $\ln{3} \approx 1.1$, we can write
\begin{equation}\label{R14}
\left(\frac{d\phi}{dt}\right)_{SQ} \approx \pm \frac{1}{3\sqrt{3}\,M}\,\left(1 - 0.15\, \frac{Q}{M^3}\right)\,.
\end{equation}

\subsubsection{Decay of the $SQ$ Light Ring}

The equatorial paths of the perturbed rays of the light ring are null; hence, $ds_{SQ}^2 = 0$ with $\theta = \pi/2$; furthermore, with $dr/dt = \epsilon r_0 f'$, $r_0 = M (3-q)$ and $d\phi/dt = \omega_{\pm} (1+\epsilon g')$, we have to linear order in $q$ and $\epsilon$,  
\begin{align}\label{M15}
\frac{1}{r^3}\,\left[r -2M(1-q)+ 2qr\mathbb{A}\ln{\mathbb{A}}\right] = \omega^2_{\pm} (1+ 2\epsilon g')\,.
\end{align} 
Using $\mathbb{A} = 1-2M/r$, the left-hand side of Eq.~\eqref{M15} reduces, after some algebra, to $\omega^2_{\pm}$; therefore, $g'=0$. Thus $g$ is a constant that must be zero in accordance with the boundary condition~\eqref{M7}. It follows that $\omega^0_{SQ} =  \pm j\,\omega_{\pm}$. It remains to determine $f(t)$ and $h(t)$ using the geodesic equation.

The null geodesic equation for the temporal coordinate can be written as
\begin{align}\label{M16}
\frac{d^2t}{d\lambda^2} + 2 \Gamma^t_{tr} \,\frac{dt}{d\lambda}\,\frac{dr}{d\lambda} = 0\,,
\end{align} 
which reduces to 
\begin{equation}\label{M17}
\epsilon \,h'' = 2\epsilon r \Gamma^t_{tr}\,f' +O(\epsilon^2)\,,
\end{equation}
where $\Gamma^{t}_{tr}$, given by Eq.~\eqref{D1} of Appendix D, can be evaluated along the unperturbed orbit.
It follows that 
\begin{equation}\label{M18}
h'' =\frac{2M}{r_0\,\mathbb{A}_0}\,\left(1-q\,\frac{2 M}{r_0\,\mathbb{A}_0}\right)\,f'\,,
\end{equation}
where $\mathbb{A}_0 = 1-2M/r_0$ and $r_0\,\mathbb{A}_0 = M(1-q)$. In this way, we find
\begin{equation}\label{M19}
h'' = 2 (1-q) f'\,.
\end{equation}

The null geodesic equation for the radial coordinate reduces, after neglecting terms of order $\epsilon^2$, to
\begin{align}\label{M20}
\frac{d^2r}{d\lambda^2} +  \Gamma^r_{tt} \,\left(\frac{dt}{d\lambda}\right)^2 +  \Gamma^r_{\phi \phi} \,\left(\frac{d\phi}{d\lambda}\right)^2 = 0\,,
\end{align} 
or, more explicitly, 
\begin{align}\label{M21}
\epsilon r_0\,f'' +  (\Gamma^r_{tt} +  \Gamma^r_{\phi \phi}\,\omega^2_{\pm})\,(1-2\epsilon h') = 0\,,
\end{align} 
where the connection coefficients are given in Appendix D. After some algebra, we find
\begin{align}\label{M22}
27 M\,\Gamma^r_{tt} = 1+2q[1-\epsilon f+ \ln(4/9)]\,  
\end{align} 
and 
\begin{align}\label{M23}
 -\frac{1}{M}\, \Gamma^r_{\phi \phi} = 1+2q\ln(4/3) +3[1+q(-1 + 2\ln(4/3)]\epsilon f\,.
\end{align} 
The end result is that $h'$ cancels out and Eq.~\eqref{M21} reduces to 
\begin{align}\label{M24}
 f'' - \frac{1}{27M^2}\,[1 +2q(1 + 2\ln(2/3))] f = 0\,.
\end{align} 

The null geodesic equation for the polar $\theta$ coordinate is automatically satisfied, since, among other things, $\sin{2\theta} = 0$.  On the other hand, the null geodesic equation for the azimuthal $\phi$ coordinate reduces to
\begin{align}\label{M25}
\frac{d^2\phi}{d\lambda^2} + 2 \Gamma^{\phi}_{r\phi} \,\frac{dr}{d\lambda}\,\frac{d\phi}{d\lambda} = 0\,,
\end{align} 
where $r^2\,\Gamma^{\phi}_{r\phi} = r -qM/\mathbb{A}$, see Appendix D. It follows that $h'' = 2 (1-q) f'$, just as in Eq.~\eqref{M19}. Using the boundary conditions~\eqref{M7}, it is now straightforward to solve Eqs.~\eqref{M19} and~\eqref{M24} and the final results are
\begin{align}\label{M26}
{}&f(t) = \beta_{SQ} \sinh(\gamma_{SQ}\, t)\,, \qquad g(t) = 0\,, \nonumber   \\
{}& h(t) =2\frac{(1-q)\beta_{SQ}}{\gamma_{SQ}}\left[\cosh(\gamma_{SQ}\,t)-1\right] + C_{SQ}\, t\,, 
\end{align} 
where $\beta_{SQ}\ne 0$ and $C_{SQ}$ are simply integration constants and $\gamma_{SQ} > 0$ is given by
\begin{align}\label{M27}
 \gamma_{SQ}^2 = \frac{1+2q\,[1+ 2 \ln(2/3)]}{27M^2}\,.
\end{align} 
It turns out that the QNM frequencies are essentially independent of $\beta_{SQ}$ and $C_{SQ}$; therefore, one can generally set $\beta_{SQ} = 1$ and $C_{SQ} = 0$. 

The last step involves the calculation of the damping rate using Eq.~\eqref{M11}. From the $SQ$-metric~\eqref{R6}, we get 
\begin{align}\label{M28}
 \sqrt{-g_{SQ}}  =  \left(1-q\ln{\frac{\mathbb{B}^2}{\mathbb{A}}}\right)\,r^2\,\sin \theta.
\end{align} 
For the case of the perturbed circular orbit under consideration here, we have
\begin{align}\label{M29}
 \sqrt{-g_{SQ}}  =  9M^2\left[1+q\left(\ln{\frac{27}{16}}-\frac{2}{3}\right)\right]\,(1+2\epsilon f).
\end{align} 
Thus Eq.~\eqref{M11} can be written as
\begin{equation}\label{M30}
\frac{1}{\rho_n}\,\frac{d\rho_n}{d\lambda} = -\epsilon \left[\frac{\partial(2f-h')}{\partial t} + \frac{\partial (r_0\,f')}{\partial r} + \omega_{\pm}\,\frac{\partial(2f-h')}{\partial \phi}\right]\,,
\end{equation}
which is evaluated \emph{along the perturbed congruence}. It follows from $dr/dt = \epsilon\, r_0\, f'$ that 
\begin{equation}\label{M31}
\frac{1}{\rho_n}\,\frac{d\rho_n}{dt} = - \frac{f''}{f'} + O(\epsilon)\,.
\end{equation}
Thus ignoring terms of order $\epsilon$, we find 
\begin{equation}\label{M32}
\rho_n(t)  = \rho_n(0) \frac{1}{\cosh(\gamma_{SQ}t)}\,.
\end{equation}
We can write this result for $t > 0$ as
\begin{equation}\label{M33}
\rho_n(t)  = 2\rho_n(0) \left(e^{-\gamma_{SQ}t} - e^{- 3\gamma_{SQ}t} + e^{- 5\gamma_{SQ}t} - \cdots\right)\,.
\end{equation}
This means that the imaginary parts of the QNM frequencies (corresponding to the damping rates of outgoing waves) are given by 
\begin{equation}\label{M34}
\Gamma_{SQ}  = \gamma_{SQ} \left( n + \frac{1}{2} \right)\,, \qquad n = 0, 1, 2, 3, \cdots\,.
\end{equation}
Finally, we can write
\begin{equation}\label{M35}
\omega_{QNM} = \omega^0_{SQ} + i \Gamma_{SQ}  = \pm \,j \omega_{\pm} + i\gamma_{SQ} \left( n + \frac{1}{2} \right)\,, \qquad n = 0, 1, 2, 3, \cdots\,,
\end{equation}
where 
\begin{align}\label{M36}
\omega_{\pm} = \pm \frac{1-q(\ln{3} - 1)}{3\sqrt{3}M}\,,\qquad  \gamma_{SQ} = \frac{1+q\,[1+ 2 \ln(2/3)]}{3\sqrt{3}M}\,,\qquad q = \frac{3Q}{2M^3}\,.
\end{align}

\subsection{QNMs of the Static $HT$ Spacetime}

\subsubsection{Light Ring for Static $HT$ Spacetime}

The static $HT$-metric is given by Eqs.~\eqref{H1}--\eqref{H3}, where $\mathcal{F}$ and $\mathcal{G}$ for $\theta = \pi/2$, $\hat{q} := 5Q/(8M^3)$ and $x = -1 + r/M$ are given by
\begin{equation}\label{M38}
\mathcal{F} = \mathbb{A}\,[1 - \hat{q}\, \mathcal{Q}_2^2(x)]\,,\qquad \mathcal{G} = 1 -  \hat{q}\,\left[\frac{2M}{r\sqrt{\mathbb{A}}}\,\mathcal{Q}_2^1(x)-\mathcal{Q}_2^2(x)\right]\,.
\end{equation}
The unstable null circular orbit of radius $r_0$ and frequency $\omega_{\pm}$ in the equatorial plane of the static $HT$-spacetime constitutes the light ring in this case.  

To determine $r_0$ , we find for the analog of Eq.~\eqref{R8} in this case
\begin{equation}\label{H14}
\left(\frac{d\phi}{dt}\right)^2 = \frac{\mathbb{A}}{r^2}\left [1 - \hat{q}\,(-16+15\ln{3})\right]\,,
\end{equation}
where for $r_{Sch} = 3M$, $x=-1+r_{Sch}/M = 2$ and hence,
\begin{equation}\label{H15}
\mathcal{Q}_2^1(2) = \sqrt{3}\, ( -3 \ln{3} +10/3)\,, \qquad \mathcal{Q}_2^2(2) = 4.5 \ln{3} -14/3\,.
\end{equation}
Similarly, for the analog of Eq.~\eqref{R9} we get
\begin{equation}\label{H16}
\left(\frac{d\phi}{dt}\right)^2 = \frac{M}{r^3}\left[1 + \frac{3}{2}\,\hat{q}\,(28-25\ln{3})\right]\,.
\end{equation}
Equating Eq.~\eqref{H14} with Eq.~\eqref{H16}, we find in the static $HT$ case
\begin{equation}\label{H17}
r_0 = 3M (1+\alpha\, \hat{q})\,, \qquad  \alpha := \frac{1}{6}(52-45\ln{3})\,.
\end{equation}
Finally, using this result, we have 
\begin{equation}\label{H18}
\frac{d\phi}{dt} = \omega_{\pm} = \pm \frac{1}{3\sqrt{3}\,M}\,\left [1 - \frac{1}{2}\hat{q}\,(-16+15\ln{3})\right]\,,
\end{equation}
or, with $\ln{3} \approx 1.1$, we get approximately the same result as in the case of the $SQ$-metric
\begin{equation}\label{H19}
\left(\frac{d\phi}{dt}\right)_{HT} \approx \pm \frac{1}{3\sqrt{3}\,M}\,\left(1 - 0.15\, \frac{Q}{M^3}\right)\,.
\end{equation}

\subsubsection{Decay of the Static $HT$ Light Ring}

Let us perturb the null circular equatorial orbit and consider the null path equation $g_{\mu \nu} dx^\mu dx^\nu = 0$; in this case, we find
\begin{equation}\label{M40}
\mathcal{F} = \frac{1}{\mathcal{F}}\,\left(\frac{dr}{dt}\right)^2 +  \mathcal{G}\,r^2\,\left(\frac{d\phi}{dt}\right)^2\,,
\end{equation}
where $dr/dt = \epsilon r_0 f'$ and $d\phi/dt = \omega_{\pm} ( 1+ \epsilon \, g')$, so that to linear order in $\epsilon$,
\begin{equation}\label{M41}
\frac{\mathcal{F}}{ \mathcal{G}\,r^2} = \frac{\mathbb{A}}{r^2}\left[1 +  \hat{q}\,\left(\frac{2M}{r\sqrt{\mathbb{A}}}\,\mathcal{Q}_2^1(x)- 2\mathcal{Q}_2^2(x)\right)\right] =  \omega^2_{\pm} ( 1+ 2\epsilon\, g')\,.
\end{equation}
We find after some algebra that 
\begin{equation}\label{M42}
\frac{\mathbb{A}}{r^2} = \frac{1-6\alpha \hat{q} \epsilon f}{27M^2}\,, \quad 1 +  \hat{q}\,\left(\frac{2M}{r\sqrt{\mathbb{A}}}\,\mathcal{Q}_2^1(x)- 2\mathcal{Q}_2^2(x)\right)  = 1+  \hat{q} \,(16-15\ln{3}) + 6\alpha \hat{q} \epsilon f\,.
\end{equation}
It follows from these results that $g'=0$; moreover,  from the boundary condition $g(0) = 0$, we conclude that $g(t) = 0$. Therefore, $\omega^0_{HT} = \pm j\,\omega_{\pm}$.

Next, we consider the geodesic equation for the temporal coordinate $t$, which reduces in this case to Eq.~\eqref{M16}, but with $\Gamma^t_{tr}$ given by Eq.~\eqref{D7} with $J=0$. Therefore, 
\begin{equation}\label{M43}
\epsilon\,h'' = 6M (1+\alpha\, \hat{q})\,\Gamma^t_{tr} \epsilon  f'\,.
\end{equation}
For  our present purposes, it is only necessary to compute $\Gamma^t_{tr}$ along the \emph{unperturbed} orbit and we find
\begin{align}\label{M45}
3M\,\Gamma^t_{tr}(\epsilon = 0)=1 - \hat{q} (24-21\ln{3})\,.
\end{align}
Therefore, the result of the temporal null geodesic equation is
\begin{align}\label{M46}
h'' = 2\left[1 - \frac{1}{6}\hat{q}\, (92-81\ln{3})\right]\, f'\,.
\end{align}

For the radial geodesic equation, Eqs.~\eqref{M20} and~\eqref{M21} are valid for the Hartle-Thorne case as well, except that we must now calculate $\Gamma^r_{tt}$
and $\Gamma^r_{\phi \phi}$ for the static $HT$ case using Appendix D. The results are 
\begin{align}\label{M49}
\Gamma^r_{tt}=\frac{1}{27M}\,\left[1+2\hat{q}\,(10-9\ln{3})+\frac{3}{2}\hat{q}\,(28-27\ln{3})\epsilon f\right]\,,
\end{align}
\begin{align}\label{M50}
-\frac{1}{M}\,\Gamma^r_{\phi \phi} = 1+\hat{q}\,(4-3\ln{3}) + [ 3 - 4\hat{q}\,(19-18\ln{3})] \epsilon f\,.
\end{align}
The end result for the radial null geodesic equation is that $h'$ drops out, as before, and we find
\begin{align}\label{M51}
 f'' - \frac{1}{27M^2}\,\left[1 + 2\hat{q}\,(-16 + 15\ln{3})\right] f = 0\,.
\end{align}   

The null geodesic equation for the $\theta$ coordinate is satisfied for our perturbed equatorial orbit. The corresponding equation for the $\phi$ coordinate reduces to Eq.~\eqref{M25}, namely,
\begin{align}\label{M52}
\epsilon\,h'' = 2 \Gamma^\phi_{r \phi}\,\epsilon r_0 f'\,,
\end{align} 
except that $\Gamma^\phi_{r \phi}$ is now for the static Hartle-Thorne case, see Appendix D. 
It is sufficient to compute this quantity along the \emph{unperturbed} circular orbit and we find 
\begin{align}\label{M54}
\Gamma^\phi_{r\phi} (\epsilon = 0) = \frac{1}{3M}\,\left[1 + 3\hat{q}\,(-8 + 7\ln{3})\right]\,.
\end{align}
This result, when inserted in Eq.~\eqref{M52}, helps us recover Eq.~\eqref{M46}. 

The solutions of the orbital perturbation equations are thus
\begin{align}\label{M55}
f(t) = \beta_{HT} \sinh(\gamma_{HT}\, t)\,, \qquad g(t) = 0\,,   
\end{align} 
\begin{equation}\label{M56}
h(t) = 2\left[1 - \frac{1}{6}\hat{q}\, (92-81\ln{3})\right]\,\frac{\beta_{HT}}{\gamma_{HT}}\left[\cosh(\gamma_{HT}\,t)-1\right] + C_{HT}\, t\,. 
\end{equation}
Here, $\beta_{HT}\ne 0$ and $C_{HT}$ are, as before, integration constants and for our present purposes we can simply set $\beta_{HT} = 1$ and $C_{HT} = 0$.  Furthermore,$\gamma_{HT} > 0$ is given by
\begin{align}\label{M57}
 \gamma_{HT} = \frac{1 +  \hat{q}\,(-16 + 15\ln{3})}{3\sqrt{3}M}\,.
\end{align} 

Finally, from the $HT$-metric~\eqref{H1} we get 
\begin{align}\label{M58}
 \sqrt{-g_{HT}}  =  \mathcal{G}\,r^2\,\sin \theta.
\end{align} 
For the perturbed orbit in the equatorial plane of the static Hartle-Thorne spacetime, we have
\begin{align}\label{M59}
 \sqrt{-g_{HT}}  =  9M^2\left[1-\frac{3}{2}\hat{q}\,(-4+3\ln{3}) + 2 \left(1+ \frac{1}{3}\hat{q}\,(-28 + 27\ln{3})\right)\,\epsilon f \right].
\end{align} 
The rate of decay of the density of null rays is given by 
\begin{align}\label{M60}
\frac{1}{\rho_n}\,\frac{d\rho_n}{d\lambda} ={}& -\frac{1}{\sqrt{-g_{HT}}}\,\frac{\partial}{\partial t}(\sqrt{-g_{HT}}\,(1-\epsilon\,h'))-\frac{1}{\sqrt{-g_{HT}}}\,\frac{\partial}{\partial r}(\sqrt{-g_{HT}}\,\epsilon\,r_0\,f') \nonumber  \\
{}& -\frac{\omega_{\pm}}{\sqrt{-g_{HT}}}\,\frac{\partial}{\partial \phi}(\sqrt{-g_{HT}}\,(1-\epsilon\,h'))\,,
\end{align}
which can be written as
\begin{equation}\label{M61}
\frac{1}{\rho_n}\,\frac{d\rho_n}{d\lambda} = -\frac{\partial(\epsilon\,r_0\,f')}{\partial r} + O(\epsilon)\,.
\end{equation}
Next, we note that, as before, along the congruence we have
\begin{equation}\label{M62}
\frac{\partial(\epsilon\,r_0\,f')}{\partial r} = \frac{d(\epsilon\,r_0\,f')}{dt}\,\left(\frac{dr}{dt}\right)^{-1} = f''/f'\,,
\end{equation}
since $dr/dt = \epsilon\, r_0\, f'$. Thus,  neglecting  terms of order $\epsilon$, we find $\rho_n(t)  = \rho_n(0)/ \cosh(\gamma_{HT}t)$\,. The QNM frequencies of the static Hartle-Thorne solution are therefore
\begin{equation}\label{M63}
\omega_{QNM} = \omega^0_{HT} + i \Gamma_{HT}  = \pm \,j \omega_{\pm} + i\gamma_{HT} \left( n + \frac{1}{2} \right)\,, \qquad n = 0, 1, 2, 3, \cdots\,,
\end{equation}
where 
\begin{equation}\label{M64}
\omega_{\pm} = \pm \frac{1 - \frac{1}{2}\hat{q}\,(-16+15\ln{3})}{3\sqrt{3}\,M}\,,\qquad   \gamma_{HT} = \frac{1 +  \hat{q}\,(-16 + 15\ln{3})}{3\sqrt{3}M}\,, \qquad \hat{q} := \frac{5Q}{8M^3}\,.
\end{equation}

\subsection{Comments on Static QNMs}

In our analytic treatment (valid in the eikonal limit), we have worked to linear order in $|Q|/M^3 \ll 1$ for the sake of simplicity. It remains to see what happens to QNM frequencies in situations where there is significant oblateness and the quadrupole contribution is thus sufficiently large. 

On the physical side,  it is important to note the following circumstance: Suppose observations determine the real part of the QNM frequency to be 
\begin{equation}\label{CM1}
\omega^0 =  \frac{j}{3\sqrt{3}\,M_{Sch}}\,
\end{equation}
up to a certain angular momentum factor $j \gg 1$, where $M_{Sch}$ is the mass of the Schwarzschild black hole involved in the physical process under investigation. On the other hand, if the collapsed configuration has mass $M$ and quadrupole moment $Q$, the relation would instead be
\begin{equation}\label{CM2}
\omega^0 \approx  \frac{j}{3\sqrt{3}\,M}\,\left(1 - 0.15\, \frac{Q}{M^3}\right)\,,
\end{equation}
based on Eqs.~\eqref{R14} and~\eqref{H19} for the $SQ$ and static $HT$ spacetimes, respectively. Equations~\eqref{CM1} and~\eqref{CM2} imply
\begin{equation}\label{CM3}
\frac{M}{M_{Sch}} \approx 1 - 0.15\, \frac{Q}{M^3}\,,
\end{equation}
so that for an oblate configuration with $Q > 0$, we have 
\begin{equation}\label{CM4}
M < M_{Sch}\,.
\end{equation}
Thus, based solely on the \emph{real part} of QNM frequency in the high-frequency regime, the actual mass of the collapsed configuration would be \emph{less} than what one would conclude if the quadrupole contribution is ignored. 
However, the situation is basically different for the imaginary part of the QNM frequency, since Eqs.~\eqref{M36} and~\eqref{M64} imply
\begin{equation}\label{CM5} 
 \gamma_{SQ} \approx \gamma_{HT} \approx \frac{1 + 0.3\,Q/M^3}{3\sqrt{3}M}\,.
\end{equation}
\emph{It is remarkable that in the eikonal limit we find essentially the same QNM frequencies for both $SQ$ and static $HT$ spacetimes to linear order in the quadrupole moment.} What is the physical origin of this coincidence? Further work is necessary to clarify this situation. 

It would be interesting to see what one would find for the least-damped dominant QNMs in the presence of quadrupole moment, since those QNMs would be most relevant from an observational point of view. 

Thus far we have neglected the rotation of the black hole. In view of the essential  equivalence of our  results for the QNM frequencies of nonrotating $SQ$ and static $HT$ spacetimes, it is interesting to determine the effect of rotation of the gravitational source.  For this purpose, we choose the Hartle-Thorne spacetime due to its physical importance~\cite{HaTh}. That is, we work out the  QNM frequencies of the rotating $HT$ spacetime when  angular momentum is included in the static $HT$ spacetime to second order of approximation. This is done in the next section.

\section{QNMs of the Rotating $HT$ Spacetime}

The metric of the rotating Hartle-Thorne spacetime is given by~\cite{HaTh}
\begin{equation}\label{L1}
ds_{RHT}^2 = -\mathbb{F}_1 \,dt^2 + \frac{1}{\mathbb{F}_2}\,dr^2 +  \mathbb{G}\,r^2\,\left[d\theta^2 +\sin^2\theta\,\left(d\phi - \frac{2J}{r^3}\,dt\right)^2\right]\,,
\end{equation}
where
\begin{equation}\label{L2}
\mathbb{F}_1 = \left(1 -\frac{2M}{r}+ \frac{2J^2}{r^4}\right)\,\left[ 1 +  \frac{2J^2}{Mr^3}\left(1+\frac{M}{r}\right)P_2(y)+ 2\tilde{q}\, \mathcal{Q}_2^2(x)\,P_2(y)\right]\,,
\end{equation}
\begin{equation}\label{L3}
\mathbb{F}_2 = \left(1 -\frac{2M}{r}+ \frac{2J^2}{r^4}\right)\,\left[ 1 +  \frac{2J^2}{Mr^3}\left(1-5\,\frac{M}{r}\right)P_2(y)+ 2\tilde{q}\, \mathcal{Q}_2^2(x)\,P_2(y)\right]\,,
\end{equation}
\begin{equation}\label{L4}
\mathbb{G} = 1 -  \frac{2J^2}{Mr^3}\left(1+2\,\frac{M}{r}\right)P_2(y)+ 2\tilde{q}\,\left[\frac{2M}{\sqrt{r(r-2M)}}\,\mathcal{Q}_2^1(x)-\mathcal{Q}_2^2(x)\right]\,P_2(y)\,.
\end{equation}
Here,  $x = -1+ r/M$ and $y = \cos\theta$, as before, while $\tilde{q}$ is a new dimensionless parameter defined by 
\begin{equation}\label{L5}
\tilde{q} := \frac{5}{8}\,\frac{Q-J^2/M}{M^3}= \hat{q} - \frac{5}{8}\,\frac{J^2}{M^4}\,.
\end{equation}
The associated Legendre functions of the second kind $\mathcal{Q}_n^m$ are given in Eqs.~\eqref{H4}--\eqref{H6}. The Einstein tensor for the $HT$ spacetime vanishes  when terms proportional to $J^3$, $JQ$, $Q^2$ and higher orders are neglected.

The rotating Hartle-Thorne solution represents the exterior field of a  configuration with mass $M$, which is treated to all orders, angular momentum $J$, which is treated to second order and quadrupole moment $Q$, which is treated to linear order in perturbation. It is therefore interesting to compare this spacetime with the exterior Kerr metric given in Eq.~\eqref{E1} of Appendix E in Boyer-Lindquist coordinates $(t, r, \theta, \phi)$.  For the Kerr metric~\eqref{E1}, the quadrupole moment is given by $Q_K = J^2/M$. With this value for the quadrupole moment $Q$,  the rotating $HT$-metric coefficients reduce to 
\begin{equation}\label{L6}
\mathbb{F}_1 = \left(1 -\frac{2M}{r}+ \frac{2J^2}{r^4}\right)\,\left[ 1 +  \frac{2J^2}{Mr^3}\left(1+\frac{M}{r}\right)P_2(y)\right]\,,
\end{equation}
\begin{align}\label{L7}
{}&\mathbb{F}_2 = \left(1 -\frac{2M}{r}+ \frac{2J^2}{r^4}\right)\,\left[ 1 +  \frac{2J^2}{Mr^3}\left(1-5\,\frac{M}{r}\right)P_2(y)\right]\,, \nonumber    \\
{}& \mathbb{G} = 1 -  \frac{2J^2}{Mr^3}\left(1+2\,\frac{M}{r}\right)P_2(y)\,.
\end{align}
The resulting metric is equivalent, up to a coordinate transformation, to the Kerr metric~\eqref{E1} when terms higher than second order in angular momentum ($J = Ma$) are neglected. Indeed, starting with the Kerr metric~\eqref{E1} valid to second order in $a$, the corresponding transformation  is given by
\begin{equation}\label{L8}
r \to r - \frac{a^2}{2r}\, \left[\left(1 +\frac{2M}{r}\right)\left(1 -\frac{M}{r}\right)-\left(1 -\frac{2M}{r}\right)\left(1 +\frac{3M}{r}\right)\,\cos^2\theta\right]\,,
\end{equation}
\begin{equation}\label{L9}
\theta \to \theta - \frac{a^2}{2r}\, \left(1 +\frac{2M}{r}\right)\,\sin\theta\,\cos\theta\,.
\end{equation}
The connection coefficients for the rotating Hartle-Thorne spacetime are given in Appendix D.

\subsection{Light Ring of the Rotating $HT$ Spacetime}

We look for an orbit with $r = r_0$ and $\theta = \pi/2$. The path of the orbit is null; therefore, $ds_{RHT}^2 = 0$ implies
\begin{equation}\label{L23}
 \left(\frac{d\phi}{dt} - \frac{2J}{r^3}\right)^2 = \frac{\mathbb{F}_1}{r^2\,\mathbb{G}}\,.
\end{equation}
Furthermore, the radial geodesic equation can be written as 
\begin{equation}\label{L24}
\Gamma^r_{\phi \phi} \left(\frac{d\phi}{dt}\right)^2 + 2\Gamma^r_{t\phi}\, \frac{d\phi}{dt} + \Gamma^r_{tt} = 0\,.
\end{equation}
We assume the null orbit is given by $\bar{x}^\mu(\lambda) = (\lambda, r_0, \pi/2, \omega_{\pm}\,\lambda)$, as before.  
It is then straightforward to check that the other components of the geodesic equation are indeed satisfied.

It follows from Eq.~\eqref{L23} that along the orbit
\begin{equation}\label{L25}
 \frac{d\phi}{dt} =   \frac{2J}{r_0^3} \pm\frac{1}{r_0} \sqrt{1 - \frac{2M}{r_0}}\,\left[ 1 +  \frac{1}{2}\,\tilde{q}\,(16-15\ln3) - \frac{1}{54}\,\frac{J^2}{ M^4}\right]\,,
\end{equation}
where $\tilde{q}$ is the parameter defined in Eq.~\eqref{L5}. On the other hand, Eq.~\eqref{L24} can be written as 
\begin{equation}\label{L26}
 \frac{d\phi}{dt} =  - \frac{\Gamma^r_{t\phi}}{\Gamma^r_{\phi \phi}} \pm  \left[\left(\frac{\Gamma^r_{t\phi}}{\Gamma^r_{\phi \phi}} \right)^2 - \frac{\Gamma^r_{tt}}{\Gamma^r_{\phi \phi}}\right]^{1/2}\,.
\end{equation}
Using the connection coefficients, we find that to the order of approximation under consideration here
\begin{equation}\label{L27}
\frac{\Gamma^r_{t\phi}}{\Gamma^r_{\phi \phi}}  = \frac{J}{r_0^3}\,
\end{equation}
and
\begin{equation}\label{L28}
 \frac{d\phi}{dt} =  - \frac{J}{r_0^3} \pm \sqrt{\frac{M}{r_0^3}}\,\left[ 1 +  \frac{3}{4}\,\tilde{q}\,(28-25\ln3) + \frac{7}{54}\,\frac{J^2}{M^4}\right]\,.
\end{equation}
From Eqs.~\eqref{L25} and~\eqref{L28}, we can find $r_0$ by assuming that $r_0 - 3M$ can be  expanded in powers of the small quantities under consideration. After some algebra, the result is
\begin{equation}\label{L29}
r_0 = 3M \mp \frac{2J}{\sqrt{3}M} + \frac{1}{2}\,M\,\tilde{q}\,(52-45\ln3) - \frac{1}{27}\,\frac{J^2}{M^3}\,.
\end{equation}
Moreover, for $d\phi/dt = \omega_{\pm}$, we can substitute $r_0$ in either Eq.~\eqref{L25} or Eq.~\eqref{L28} to get
\begin{equation}\label{L30}
\omega_{\pm} =  \pm \frac{1}{3\sqrt{3}M}\,\left[1\pm \frac{2J}{3\sqrt{3}M^2} - \frac{1}{2}\,\tilde{q}\,(-16+15\ln3) + \frac{11}{54}\,\frac{J^2}{M^4}\right]\,.
\end{equation}
For $J = 0$, $\tilde{q} = \hat{q}$ and we recover our previous result given in Eq.~\eqref{H18} of Section V.

\subsection{Decay of the Light Ring}

As before, we assume that a bundle of null rays initially moving on the light ring, $\bar{x}^\mu = (t, r, \theta, \phi) = (\lambda, r_0, \frac{\pi}{2}, \omega_{\pm}\,\lambda)$, begins to decay at $t = 0$, so that the paths of the null geodesics for $t > 0$ are given by $r = r_0 [ 1 + \epsilon\,f(t)]$,  $\phi = \omega_{\pm} [t + \epsilon\,g(t)]$ and  $\lambda = t + \epsilon\,h(t)$, where $f(0) = g(0) = h(0) = 0$. The bundle's ingoing and outgoing paths are null; therefore, we have
\begin{equation}\label{L31}
r^2\,\mathbb{G}\, \left(\frac{d\phi}{dt} - \frac{2J}{r^3}\right)^2 = \mathbb{F}_1 + O(\epsilon^2)\,.
\end{equation}
Next, we use the perturbed path components in Eq.~\eqref{L31} and note that
\begin{equation}\label{L32}
\frac{\mathbb{F}_1}{r^2\,\mathbb{G}} = \frac{1}{27M^2}\,\left[ 1 + \tilde{q}\,(16-15\ln3) \pm \frac{4J}{\sqrt{3}M^2}\,\epsilon f-\frac{J^2}{27 M^4}(13-72\epsilon f)\right]   + O(\epsilon^2)\,,
\end{equation}
where $\tilde{q}$ has been defined by Eq.~\eqref{L5}. 
Finally,  employing Eqs.~\eqref{L29}--\eqref{L32}, we find after much algebra that $g'(t) = 0$. Thus, as before, $g(t) = 0$ follows from the boundary condition that $g(0) = 0$. Therefore, for the rotating Hartle-Thorne case we have $\omega^0_{RHT} = \pm j\,\omega_{\pm}$.

The temporal component of the geodesic equation can be written as
\begin{equation}\label{L33}
\epsilon\,h'' = 2 r_0\,( \Gamma^t_{tr} + \Gamma^t_{r\phi}\,\omega_{\pm})\,\epsilon\,f'  +O(\epsilon^2)\,.
\end{equation}
Along the \emph{unperturbed} orbit, we find
\begin{equation}\label{L34}
 \Gamma^t_{tr}(\epsilon = 0) = \frac{1}{3M}\,\left[ 1\pm \frac{8J}{3\sqrt{3}M^2} + 3\,\tilde{q}\,(-8 + 7\ln3) + \frac{37}{18}\,\frac{J^2}{M^4}\right]\,
\end{equation}
and 
\begin{equation}\label{L35}
\omega_{\pm}\, \Gamma^t_{r\phi}(\epsilon = 0) = \frac{1}{3M}\,\left[\mp \frac{J}{\sqrt{3}M^2}- \frac{10}{9}\,\frac{J^2}{M^4}\right]\,.
\end{equation}
Next, employing Eq.~\eqref{L29} for $r_0$, Eq.~\eqref{L33} takes the form
\begin{equation}\label{L36}
 h'' = 2\,\left[ 1\pm \frac{J}{\sqrt{3}M^2} -  \frac{1}{6}\,\tilde{q}\,(92-81\ln3)  + \frac{91}{162}\,\frac{J^2}{M^4}\right]\,f'\,.
\end{equation}

The radial component of the geodesic equation yields
\begin{equation}\label{L37}
\epsilon\,r_0\,f'' + ( \Gamma^r_{tt} + 2\Gamma^r_{t\phi}\,\omega_{\pm} + \Gamma^r_{\phi \phi}\,\omega_{\pm}^2 )\,(1-2\epsilon\,h' ) +O(\epsilon^2) = 0\,.
\end{equation}
Employing the Christoffel symbols given in Appendix D, after much algebra Eq.~\eqref{L37} takes the form
\begin{align}\label{L38}
 f'' - \frac{1}{27M^2}\,\left[1 + 2\,\tilde{q}\,(-16 + 15\ln{3}) - \frac{4J^2}{27M^4}\right] f = 0\,.
\end{align} 
When $J = 0$, $\tilde{q} = \hat{q}$ and our result reduces to Eq.~\eqref{M51}.  

As before, the null geodesic equation for the $\theta$ coordinate is satisfied and the corresponding equation for the $\phi$ coordinate reduces to
\begin{align}\label{L39}
\epsilon\,\omega_{\pm}\,h'' = 2 (\Gamma^\phi_{t r} + \Gamma^\phi_{r \phi}\,\omega_{\pm})\,\epsilon r_0 f'\,.
\end{align}
From $\Gamma^\phi_{t r} = J/(r^4 \mathbb{A})$, we find at the order of approximation under consideration
\begin{align}\label{L40}
\Gamma^\phi_{t r}(\epsilon = 0) = \frac{J}{27M^4}\left( 1 \pm \frac{4J}{\sqrt{3}M^2}\right)\,.
\end{align}
Moreover, $\Gamma^\phi_{r \phi}$ in the rotating Hartle-Thorne case is given by
\begin{align}\label{L41}
\Gamma^\phi_{r\phi} (\epsilon = 0) = \frac{1}{3M}\,\left[1 \pm \frac{2J}{3\sqrt{3}M^2}+ 3\,\tilde{q}\,(-8 + 7\ln{3}) - \frac{J^2}{6M^4}\right]\,.
\end{align}
With these results, Eq.~\eqref{L39} reduces to Eq.~\eqref{L36}. 

The orbital perturbations follow from the solutions of the null geodesic equation, namely,  
\begin{align}\label{L42}
f(t) = \beta_{RHT} \sinh(\gamma_{RHT}\, t)\,, \qquad g(t) = 0\,,   
\end{align} 
\begin{align}\label{L43}
h(t) ={}& 2\left[ 1\pm \frac{J}{\sqrt{3}M^2} -  \frac{1}{6}\,\tilde{q}\,(92-81\ln3)  + \frac{91}{162}\,\frac{J^2}{M^4}\right]\,\frac{\beta_{RHT}}{\gamma_{RHT}}\left[\cosh(\gamma_{RHT}\,t)-1\right]  \nonumber  \\
& + C_{RHT}\, t\,. 
\end{align}
As before, the integration constants $\beta_{RHT}\ne 0$ and $C_{RHT}$ can be set equal to unity and zero, respectively.  Furthermore,$\gamma_{RHT} > 0$ is given by
\begin{align}\label{L44}
 \gamma_{RHT} = \frac{1 +  \tilde{q}\,(-16 + 15\ln{3}) - 2J^2/27M^4}{3\sqrt{3}M}\,.
\end{align} 

Regarding the calculation of the relative rate of decay of the density of null rays along the light ring, let us first note that for the rotating $HT$-metric~\eqref{L1},
\begin{align}\label{L45}
 \sqrt{-g_{RHT}}  = \sqrt{\frac{\mathbb{F}_1}{\mathbb{F}_2}}\, \mathbb{G}\,r^2\,\sin \theta.
\end{align} 
Along the perturbed orbit we have,
\begin{align}\label{L46}
\frac{1}{9M^2}\, \sqrt{-g_{RHT}}  = {}&1 \mp \frac{4J}{3\sqrt{3}M^2} -\frac{3}{2}\,\tilde{q}\,(-4+3\ln{3}) + \frac{4J^2}{27M^4} \nonumber  \\
{}&+ 2 \left[1\mp \frac{4J}{3\sqrt{3}M^2}+ \frac{1}{3}\,\tilde{q}\,(-28 + 27\ln{3})+ \frac{19J^2}{162M^4} \right]\,\epsilon f .
\end{align} 
The relative rate of decay of the density of null rays can now be calculated as in the static Hartle-Thorne case and the result is
\begin{align}\label{L47}
\rho_n(t)  = \rho_n(0)/ \cosh(\gamma_{RHT}\,t)\,,
\end{align}
when we neglect terms of order $\epsilon$.
Finally, the QNM frequencies of the rotating Hartle-Thorne solution via the light-ring method are given by
\begin{align}\label{L48}
\omega^0_{RHT} + i \Gamma_{RHT}  = {}& \frac{j}{3\sqrt{3}M}\,\left[1 \pm \frac{2J}{3\sqrt{3}M^2} - \frac{1}{2}\,\tilde{q}\,(-16 +15\ln3) + \frac{11}{54}\,\frac{J^2}{M^4}\right]  \nonumber   \\
&+ i\,\frac{n + \frac{1}{2}}{3\sqrt{3}M}\left[1 +  \tilde{q}\,(-16 + 15\ln{3}) - \frac{2}{27}\,\frac{J^2}{M^4}\right]\,, \quad n = 0, 1, 2, \cdots\,,
\end{align}
where $\tilde{q} = 5(Q - J^2/M)/(8M^3)$.
The QNM frequencies of the rotating Hartle-Thorne solution can be compared with those of the Kerr metric to second order in $J/M^2$: For $Q_K = J^2/M$, $\tilde{q} = 0$ and the QNM frequencies of the rotating Hartle-Thorne solution coincide with those of the Kerr solution given in Appendix E. 

\section{Discussion}

In this paper, we have investigated collapsed configurations with mass $M$, angular momentum $J$ and quadrupole moment $Q$.  To linear order in $Q$ and second order in $J$,  the QNM frequencies of one such system, namely, the stationary exterior Hartle-Thorne spacetime have been analytically calculated in the eikonal limit using the light-ring method. That is, for massless field perturbations of spin $s$, $s = 0, 1, 2$,  with parameters $(\omega, j, \mu)$ such that $M\omega \gg 1$, $j \gg 1$ and $\mu = \pm j$, we have computed the ringdown frequencies, given by Eq.~\eqref{L48},  that reduce to those of the Kerr black hole in appropriate limits.  The deviation of the QNM frequencies of the rotating Hartle-Thorne system from those of the corresponding Kerr black hole can be measured in terms of the dimensionless parameter $\tilde{q}= 5(Q - J^2/M)/(8M^3)$. When the magnitude of this parameter is sufficiently small, a generalized black hole with classical quadrupole moment cannot be observationally distinguished from a Kerr black hole through their QNM frequencies. 

The determination of the QNM frequencies of the rotating $SQ$ spacetime requires further investigation.  Moreover, astrophysical (i.e., generalized)  black holes under consideration in the present work are expected to have accretion disks and associated electromagnetic fields, which we have neglected for the sake of simplicity. Our results may be relevant for considerations related to static quadrupolar perturbations generated by \emph{tidal} interactions~\cite{Hart, Konoplya:2012vh}.

\appendix

\section{Legendre Functions} 

\subsection{$P_n(z)$ and $Q_n(z)$ for $z\in [-1,1]$}

We recall that Legendre polynomials are given by
\begin{equation}\label{a1}
P_0(z) = 1\,,\qquad  P_1(z) = z\,, \qquad P_2(z) = \frac{1}{2} (3z^2-1)\,, \qquad P_3(z) = (5z^3-3z)/2\,,
\end{equation}
etc., where $P_n(1) = 1$. Moreover, the first three Legendre functions of the second kind $Q_n(z)$ are given by~\cite{AS}
\begin{align}\label{a2}
Q_0(z) = {}&\frac{1}{2} \ln\left(\frac{1+z}{1-z}\right)\,,\quad  Q_1(z) =  \frac{z}{2} \ln\left(\frac{1+z}{1-z}\right)-1\,,   \nonumber   \\
{}& Q_2(z) = \frac{3z^2-1}{4} \ln\left(\frac{1+z}{1-z}\right) - \frac{3}{2}z\,.
\end{align}

\subsection{$\mathcal{P}_n(x)$ and $\mathcal{Q}_n(x)$ for $x\in [1, \infty)$}

For $x \ge 1$, the Legendre function of the second kind can be expressed as
\begin{align}\label{A1}
\mathcal{Q}_n(x)=\frac{1}{2}\mathcal{P}_n (x)\ln\,\left(\frac{x+1}{x-1} \right) -\dfrac{2n-1}{n}\mathcal{P}_{n-1}(x)-\dfrac{2n-5}{3(n-1)}\mathcal{P}_{n-3}(x) - \cdots\,,
\end{align}
where $\mathcal{P}_n (x)$ is indeed the same as the Legendre polynomial $P_n(x)$ but for $x \ge 1$.  Hence, 
\begin{align}
\label{A2}
\mathcal{Q}_0(x) = {} & \frac{1}{2} \ln\left(\frac{x+1}{x-1}\right)\,,\quad  \mathcal{Q}_1(x) =  \frac{x}{2} \ln\left(\frac{x+1}{x-1}\right)-1\,, \\
\label{A3}                {}& \mathcal{Q}_2(x) = \frac{3x^2-1}{4} \ln\left(\frac{x+1}{x-1}\right) - \frac{3x}{2}\,,
\end{align}
etc., where $\mathcal{Q}_n(\infty) = 0$. Moreover, for $x\in (-\infty, -1]$, 
\begin{equation}\label{A4}
\mathcal{P}_n (x) = (-1)^{n}\,\mathcal{P}_n (-x)\,, \qquad \mathcal{Q}_n (x) = (-1)^{n+1}\,\mathcal{Q}_n (-x)\,.
\end{equation}

Finally, in connection with the \emph{associated} Legendre functions of the second kind defined in Eq.~\eqref{H4}, we have the recurrence relations
\begin{align}
\label{A5} &(2n+1)\,x\, \mathcal{Q}_n^m(x)=\left(n-m+1 \right) \mathcal{Q}_{n+1}^m(x)+\left( n+m\right)\mathcal{Q}_{n-1}^m(x)\,,\\
\label{A6} &\left(x^2-1 \right)\frac{d\mathcal{Q}_n^m(x)}{dx} =n x\mathcal{Q}_n^m(x)-\left(n+m \right) \mathcal{Q}_{n-1}^m(x)\,,
\end{align}
etc.

\section{Curvature Invariants of the $\delta$-Metric}

A Ricci-flat solution in GR has four algebraically independent scalar polynomial curvature invariants  given by
\begin{equation}\label{B1}
I_1 = R_{\mu \nu \rho \sigma}\,R^{\mu \nu \rho \sigma} - i R_{\mu \nu \rho \sigma}\,^{*}R^{\mu \nu \rho \sigma}\,
\end{equation}  
and
\begin{equation}\label{B2}
I_2 = R_{\mu \nu \rho \sigma}\,R^{\rho \sigma \alpha \beta}\,R_{\alpha \beta}{}^{\mu \nu} + i R_{\mu \nu \rho \sigma}\,R^{\rho \sigma \alpha \beta}\,^{*}R_{\alpha \beta}{}^{\mu \nu}\,.
\end{equation}
If the spacetime under consideration is algebraically special, we have $I_1^3 = 12\,I_2^2$.
Here, we define the dual curvature tensor $^{*}R_{\mu \nu \rho \sigma}$ via
\begin{equation}\label{B3}
^{*}R_{\mu \nu \rho \sigma} = \frac{1}{2}\,e_{\mu \nu \alpha \beta}\,R^{\alpha \beta}{}_{\rho \sigma}\,,
\end{equation}
where $e_{\mu \nu \rho \sigma}$ denotes the alternating tensor given by
$e_{\mu \nu \rho \sigma} = \sqrt{-g}\,\epsilon_{\mu \nu \rho \sigma}$ and $\epsilon_{\mu \nu \rho \sigma}$ is the  totally antisymmetric symbol with $\epsilon_{0123} = 1$. Note that in a Ricci-flat spacetime, the right and left duals of the (Weyl) curvature tensor are equal. 

It turns out that $I_1$ and $I_2$ for the $\delta$-metric are both real and can be expressed as
\begin{equation}\label{B4}
 I_1 = \frac{16 \delta^2 m^6}{r^{10}}\,\frac{\mathbb{B}^{2\delta^2-3}}{\mathbb{A}^{2(\delta^2-\delta +1)}}\,\mathcal{N}_1\,, \qquad
 I_2 = \frac{48 \delta^3 m^8}{r^{14}}\,\frac{\mathbb{B}^{3\delta^2-4}}{\mathbb{A}^{3(\delta^2-\delta +1)}}\,(X-\delta - 1)\,\mathcal{N}_2\,,
\end{equation}
where  $\mathcal{N}_1$ and $\mathcal{N}_2$ are dimensionless, $X := r/m$, 
\begin{align}\label{B5}
\mathcal{N}_1 = {} &  3X^4 -6(\delta +2) X^3 +3[(\delta +1)(\delta+5) +\delta^2 \sin^2\theta]X^2\\
\nonumber  & -3(\delta+1)^2 (2 + \delta\,\sin^2\theta)X + (\delta +1)^2(\delta ^2+\delta +1)\,\sin^2\theta\,,
\end{align}
\begin{align}\label{B6}
\mathcal{N}_2 = {} &  2X^4 -4(\delta +2) X^3 +[2(\delta +1)(\delta+5) + (3\delta^2-1) \sin^2\theta]X^2\\
\nonumber  & -(\delta+1)[4(\delta+1) + (3\delta^2 +3\delta -2)\sin^2\theta]X + \delta (\delta +1)^3\,\sin^2\theta\,.
\end{align}
Our result for the Kretschmann scalar (i.e., the real part of $I_1$) agrees with that given in Ref.~\cite{Boshkayev:2015jaa}. Moreover, for $\delta = 0$, $I_1 = I_2 = 0$ and the resulting $\delta$-metric is flat, since it can be obtained from the Minkowski metric in spatial cylindrical  coordinates via the coordinate transformations $\rho = \sqrt{r(r-2m)}\,\sin\theta$ and $z = (r-m) \,\cos\theta$.

Curvature singularities of the $\delta$-metric occur at $r=0$ and $r=2m$. Furthermore, for $0 < \delta < \sqrt{3/2}$ (excluding $\delta = 1$), curvature singularities occur at $r_1=m(1+\cos \theta)$ and $r_2=m(1-\cos \theta)$, where $\mathbb{B}=0$. For further discussion, see Ref.~\cite{Papadopoulos:1981wr}.

Let us now compute $I_1^3 - 12 I_2^2$ for the $\delta$-metric. The result is
\begin{equation}\label{B7}
I_1^3 - 12 I_2^2 = \frac{2^{10} m^{18} \delta^6}{r^{30}}\frac{\mathbb{B}^{3(2\delta^2-3)}}{\mathbb{A}^{6(\delta^2-\delta+1)}}\,(\delta^2-1)^2\,\sin^4 \theta\,\mathcal{N}_{3}^2\,\mathcal{N}_4\,,
\end{equation}
where $\mathcal{N}_3 = 3X^2 - 3 X (\delta + 2) + (\delta +1)(\delta +2)$ and 
\begin{equation}\label{B8}
\mathcal{N}_4 =  9X^4 -18 (\delta + 2) X^3 + 3 X^2 W_1 + 6 (\delta + 1) X W_2 + W_3\, \sin^2\theta\,.
\end{equation}
Here, 
\begin{equation}\label{B9}
W_1 =  3 (1+\delta)(5+ \delta) + (4\delta^2-1)\, \sin^2\theta\,,
\end{equation}
\begin{equation}\label{B10}
W_2 =  -3 (1+ \delta) + (1-2\delta -2\delta^2)\, \sin^2\theta\,,\qquad   W_3 = (1 + \delta)^2(1+2 \delta)^2\,.
\end{equation}
It follows that $\delta$-metric is of Petrov type I~\cite{Papadopoulos:1981wr}, since $I_1^3 - 12 I_2^2$ is in general nonzero. However, $I_1^3 - 12 I_2^2$ vanishes for $\theta = 0, \pi$, which means that $\delta$-metric is algebraically special all along the symmetry axis. This conclusion is in agreement with the result of Ref.~\cite{Papadopoulos:1981wr} that the $\delta$-metric is of Petrov type D on the axis ($\theta = 0, \pi$).
Furthermore, $\delta$-metric is degenerate as well when $\mathbb{B}=0$ (for $\delta > \sqrt{3/2}$) and at the positive roots of $\mathcal{N}_3 = 0$ and $\mathcal{N}_4 = 0$. In connection with $\mathbb{B}=0$, the metric is degenerate on $r_1=m(1+\cos \theta)$ and $r_2=m(1-\cos \theta)$. For $\delta \le 2$, the roots of $\mathcal{N}_3 = 0$ are positive and the metric is algebraically special at $r = m X$, where $2X = 2+\delta + \sqrt{(4-\delta^2)/3}$. Finally, for an oblate object ($\delta > 1$), the number of positive roots of $\mathcal{N}_4 = 0$ can be four, two or zero by Descartes' rule of signs; in fact, the explicit determination of these roots is in general straightforward but complicated and is beyond the scope of this appendix.

\section{Gravity Gradiometry Near $r = 2 m$}

In an arbitrary gravitational field expressed in admissible coordinates, it is always possible to define observers that remain at rest in space. We are interested in such \emph{static observers} in $\delta$-spacetime.  For $r > 2m$, we consider an arbitrary static observer that carries an orthonormal tetrad frame $\lambda^{\mu}{}_{\hat \alpha}$, where $\lambda^{\mu}{}_{\hat 0} = u^\mu$ is
the 4-velocity of a static observer and its local spatial frame is given by $\lambda^{\mu}{}_{\hat i}$ for $i = 1, 2, 3$. In $(t, r, \theta, \phi)$ coordinates, we have
\begin{equation}\label{C1}
\lambda_{\hat 0} = \mathbb{A}^{-\delta/2}\,\partial_t\,,\quad \lambda_{\hat 1} =  \mathbb{A}^{\delta/2}\, \left(\frac{\mathbb{A}}{\mathbb{B}}\right)^{(1-\delta^2)/2}\,\partial_r\,,
\end{equation}
\begin{equation}\label{C2}
\lambda_{\hat 2} = \frac{1}{r}\,\mathbb{A}^{(\delta - 1)/2}\,\left(\frac{\mathbb{A}}{\mathbb{B}}\right)^{(1-\delta^2)/2}\,\partial_{\theta}\,,\quad \lambda_{\hat 3} = \frac{1}{r\,\sin\theta}\,\mathbb{A}^{(\delta - 1)/2}\,\partial_{\phi}\,.
\end{equation}

The static observer is in general accelerated and measures the curvature tensor via a gravity gradiometer. The measured components of the curvature tensor are given by
\begin{equation}\label{C3}
R_{\hat \alpha \hat \beta \hat \gamma \hat \delta} = R_{\mu \nu \rho \sigma}\,\lambda^{\mu}{}_{\hat \alpha}\,
\lambda^{\nu}{}_{\hat \beta}\,\lambda^{\rho}{}_{\hat \gamma}\,\lambda^{\sigma}{}_{\hat \delta}\,,
\end{equation}
which are obtained from the projection of the Riemann curvature tensor upon the  tetrad frame of the reference observer along its world line. The Riemann curvature tensor has 20 independent components. Taking advantage of the symmetries of the Riemann tensor, this quantity can be represented by a $6\times6$ matrix $\mathcal{R} = (\mathcal{R}_{\hat{\mathcal{I}} \hat{\mathcal{J}}})$, where the indices $\mathcal{I}$ and $\mathcal{J}$  range over the set $(01,02,03,23, 31,12)$. Thus we can write
\begin{equation}
\label{C4}
\mathcal{R}=\left[
\begin{array}{cc}
\mathcal{E} & \mathcal{B}\cr
\mathcal{B^{\dagger}} & \mathcal{S}\cr
\end{array}
\right]\,,
\end{equation}
where $\mathcal{E}$  and $\mathcal{S}$ are symmetric $3\times3$ matrices and  $\mathcal{B}$ is traceless. The tidal matrix $\mathcal{E}$ represents the ``electric" components of the curvature tensor as measured by the fiducial observer, whereas $\mathcal{B}$ and $\mathcal{S}$ represent its ``magnetic" and ``spatial" components, respectively.  In Ricci-flat regions of spacetime, Eq.~\eqref{C4} simplifies, since   
$\mathcal{S} = -\mathcal{E}$, $\mathcal{E}$ is traceless and $\mathcal{B}$ is symmetric. Hence, the Weyl curvature tensor with 10 independent components is completely determined by its ``electric" and ``magnetic" components that are symmetric and traceless $3\times3$ matrices. 

The Weyl tensor of the $\delta$-metric as measured by the static observer has no gravitomagnetic components, so that $\mathcal{B} = 0$, as expected. The measured curvature tensor's nonzero gravitoelectric components are given by ($X := r/m$)
\begin{equation}\label{C5}
\mathcal{E}_{\hat1 \hat1} = -\mathcal{E}_{\hat 2 \hat 2} - \mathcal{E}_{\hat 3 \hat 3}\,,
\end{equation}
\begin{equation}\label{C6}
\mathcal{E}_{\hat 2 \hat 2}  = \frac{m^4\delta}{r^6} \frac{\mathbb{B}^{\delta^2-2}}{\mathbb{A}^{\delta^2-\delta+1}}\,\left\{X^3-(\delta + 3)\,X^2 + [2(\delta +1) +\delta^2\sin^2\theta]\,X -\delta (\delta + 1)\sin^2\theta\right\}\,,
\end{equation}
\begin{equation}\label{C7}
\mathcal{E}_{\hat 3 \hat 3}  = \frac{m^2\delta}{r^4}\frac{\mathbb{B}^{\delta^2-1}}{\mathbb{A}^{\delta^2-\delta+1}}\,(X-\delta-1)\,,
\end{equation}
\begin{equation}\label{C8}
\mathcal{E}_{\hat 1 \hat 2}  = \mathcal{E}_{\hat 2 \hat 1} = \frac{m^3\delta}{r^5}\frac{\mathbb{B}^{\delta^2-2}}{\mathbb{A}^{\delta^2-\delta+\frac{1}{2}}}\,(\delta^2-1)\sin \theta  \cos \theta\,.
\end{equation}
As expected, the off-diagonal component vanishes in the Schwarzschild case ($\delta = 1$). Moreover, as $r \to \infty$, $\mathcal{E}$ becomes proportional to $M/r^3$, where $M = m\delta$; indeed, $\mathcal{E}/(M/r^3) \sim $~diag(-2, 1, 1), as expected from Newtonian gravity. 

Let us note here the directional nature of the naked singularity at $r = 2 m$. For $\theta = 0, \pi$, the measured gravitoelectric components of the Weyl curvature vanish if $\delta >2$ as we approach $r = 2 m$ along the symmetry axis, but  diverge as $\mathbb{A}^{\delta - 2}$ if $\delta < 2$~\cite{Papadopoulos:1981wr}. However, in the equatorial plane $\theta = \pi/2$, the gravitoelectric  components always diverge as $\mathbb{A}^{-\delta^2 +\delta - 1}$.

\section{Christoffel Symbols}

\subsection{$SQ$-Metric}

The nonzero connection coefficients for the $SQ$-metric~\eqref{R6} can be obtained from the expressions given below, where $q := 3Q/(2M^3)$.
\begin{equation}
 \label{D1}\Gamma^{t}_{tr}=\frac{M}{r^2\,\mathbb{A}}\,\left(1-q\,\frac{2 M}{r\,\mathbb{A}}\right)\,, \qquad \Gamma^{r}_{tt}=\frac{M\mathbb{A}}{r^2}\,\left[1+ 2q\,\left(\frac{M}{r\,\mathbb{A}}+\ln\mathbb{B}\right)\right]\,, 
\end{equation}
\begin{equation}
 \label{D2}\Gamma^{r}_{rr}=-\frac{M}{r^2\,\mathbb{A}}\,\left[1-2q\,M\,\frac{r^2\,\mathbb{A}+ (r^2- 3\,Mr+3\,M^2)\,\sin^2\theta}{r^3\,\mathbb{A}\,\mathbb{B}}\right]\,, \quad \Gamma^{r}_{r\theta}=-\frac{q M^2}{r^2\mathbb{B}}\, \sin 2\theta\,,  
\end{equation}
\begin{equation}  
 \label{D3}\Gamma^{r}_{\theta \theta }=-r\,\mathbb{A}\,\left(1+q\,M\frac{r^2 - 2 Mr \cos^2\theta -M^2\sin^2\theta}{r^3\,\mathbb{A}\,\mathbb{B}}\right)\,,  
\end{equation} 
\begin{equation}
\label{D4}\Gamma^{r}_{\phi \phi }= - r\,\mathbb{A}\,\left[1+q\,\left(\frac{M}{r\,\mathbb{A}} + 2\,\ln\frac{\mathbb{B}}{\mathbb{A}}\right)\right]\,\sin^2\theta\,, \qquad \Gamma^{\theta }_{rr}=\frac{q\,M^2}{r^4\mathbb{A}\mathbb{B}}\,\sin 2\theta\,,   
\end{equation}
\begin{equation}
\label{D5}\Gamma^{\theta }_{r\theta}=\frac{1}{r}\,\left[1-q\,M\,\frac{r^2-2Mr(1+\sin^2\theta) + 3 M^2 \sin^2\theta}{r^3\,\mathbb{A}\,\mathbb{B}}\right]\,, \qquad \Gamma^{\theta}_{\theta \theta }=-\frac{q\,M^2}{r^2\mathbb{B}}\,\sin2\theta\,,   
\end{equation}
\begin{equation}
 \label{D6}\Gamma^{\theta }_{\phi  \phi }=-\frac{1}{2}\left(1 + 2q\,\ln\frac{\mathbb{B}}{\mathbb{A}}\right)\,\sin 2\theta\,, \qquad \Gamma^{\phi }_{r\phi }=\frac{1}{r}\left(1-q\,\frac{M}{r\,\mathbb{A}}\right)\,, \qquad \Gamma^{\phi }_{\theta \phi }=\cot\theta\,.  
\end{equation}

\subsection{Rotating $HT$-Metric}
 
The nonzero connection coefficients for the rotating Hartle-Thorne solution~\eqref{L1} can be obtained from the expressions given below. Here, $\tilde{q} = 5(Q - J^2/M)/(8M^3)$.
\begin{align}\label{D7}
\Gamma^t_{tr}=  {}&-\frac{\tilde{q}}{M^2}\left[3(r-M)\ln\mathbb{A}+\frac{2M}{r^4\mathbb{A}^2}(3r^4 -12Mr^3 + 13M^2r^2 - 2M^3r +2M^4)\right] P_2(\cos \theta)  \nonumber \\
{}& +\frac{M}{r^2 \mathbb{A}} -  \frac{J^2}{Mr^6\mathbb{A}}(3r^2+2Mr-8M^2)P_2(\cos \theta) - 2 \frac{MJ^2}{r^6\mathbb{A}^2}\,,
\end{align}
\begin{align}\label{D8}
\Gamma^t_{t\theta }={}&\frac{3\,\tilde{q}}{4M^2}\,\left[3r^2\mathbb{A}\ln\mathbb{A}+\frac{2M}{r^2\mathbb{A}}\,(r-M)(3r^2-6Mr -2M^2)\right] \sin 2\theta    \nonumber \\
{}&-\frac{3J^2}{2Mr^4}(r+M) \sin 2\theta\,,
\end{align}
\begin{align}\label{D9}
\Gamma^t_{r\phi }=-\frac{3J\sin^2\theta}{r^2 \mathbb{A}}\,,\quad \Gamma^r_{t\phi }=-\frac{J}{r^3}\,(r-2M)\, \sin^2\theta\,,\quad \Gamma^r_{\phi  \phi }=\Gamma^r_{\theta  \theta }\,\sin^2\theta\,,
\end{align}
\begin{align}\label{D10}
\Gamma^r_{tt}=  {}&-\frac{\tilde{q}}{M^2}\left[3(r+M)\mathbb{A}^2 \ln\mathbb{A}+\frac{2M}{r^4}(3r^4 - 6Mr^3-5M^2r^2 +6M^3r +6M^4)\right] P_2(\cos \theta)  \nonumber   \\
{}& +\frac{M \mathbb{A}}{r^2} + \frac{2J^2}{3r^6} (2r-M) -\frac{J^2 \mathbb{A}}{3Mr^5}\,(9r-2M)P_2(\cos \theta)\,,     
\end{align}
\begin{align}\label{D11}
\Gamma^r_{rr}= - \Gamma^t_{tr}-\frac{4J^2}{r^6\mathbb{A}}(7r-12M)\,P_2(\cos \theta) + \frac{4J^2}{r^5\mathbb{A}}\,,\quad \Gamma^r_{r\theta }= - \Gamma^t_{t\theta }-\frac{9J^2}{r^4}\,\sin 2\theta\,,
\end{align}
\begin{align}\label{D12}
\Gamma^r_{\theta  \theta }= {}&2M-r-\frac{\tilde{q}}{M^2}\left[2M(3r^2+3Mr-14M^2)+3r(r^2+2Mr-2M^2)\mathbb{A} \ln\mathbb{A}\right] P_2(\cos \theta)\,    \nonumber   \\
{}& -\frac{J^2}{Mr^4}\,(3r^2-12Mr+12M^2)P_2(\cos \theta) - \frac{2J^2}{r^3}\,,
\end{align}
\begin{align}\label{D13}
\Gamma^\theta _{tt}= \frac{\mathbb{A}}{r^2}\,\Gamma^t_{t\theta} -\frac{2J^2}{r^6}\,\sin2\theta\,, \quad \Gamma^\theta _{t\phi }=\frac{J\sin2\theta}{r^3}\,, \quad
\Gamma^\theta _{rr}= \frac{1}{r^2\mathbb{A}}\,\Gamma^t_{t\theta } +\frac{9J^2}{r^6\mathbb{A}}\,\sin2\theta\,,   
\end{align}
\begin{align}\label{D14}
\Gamma^\theta _{r\theta }= {}&\frac{1}{r}+\frac{\tilde{q}}{M^2}\left[3r\ln\mathbb{A}+\frac{2M}{r^3\mathbb{A}}\,(3r^3 - 3Mr^2-2M^2r -2M^3)\right] P_2(\cos \theta)   \nonumber   \\
{}&+\frac{J^2}{Mr^5}\,(3r+8M)P_2(\cos \theta)\,,
\end{align}
\begin{align}\label{D15}
\Gamma^\theta _{\theta  \theta }={}&-\frac{3\tilde{q}}{4M^2}\left[3(r^2-2M^2) \ln\mathbb{A}+\frac{2M}{r}\,(3r^2+3Mr-2M^2)\right] \sin2\theta  \nonumber   \\
{}&+\frac{3J^2}{2Mr^4}(r+2M)\sin2\theta\,,
\end{align}
\begin{align}\label{D16}
\Gamma^\theta _{\phi  \phi }= -\Gamma^\theta _{\theta  \theta }\,\sin^2\theta -\frac{1}{2}\,\sin2 \theta\,,\qquad  \Gamma^\phi_{\theta \phi }=\Gamma^\theta _{\theta  \theta } +\cot\theta\,,
\end{align}
\begin{align}\label{D17}
\Gamma^\phi _{tr}=\frac{J}{r^4\mathbb{A}}\,, \qquad \Gamma^\phi _{t\theta }=-\frac{2J\cot\theta}{r^3}\,,\qquad  \Gamma^\phi _{r\phi }= \Gamma^\theta _{r\theta }-\frac{6J^2}{r^5\mathbb{A}}\,\sin^2\theta\,.
\end{align}

\section{Kerr QNMs to $(a/M)^2$ order via the Light-Ring Method}

The metric of the exterior Kerr spacetime is given by~\cite{Chandra}
\begin{equation}\label{E1}
ds^2 = -dt^2+\frac{\Sigma}{\Delta}dr^2+\Sigma\, d\theta^2 +(r^2+a^2)\sin^2\theta\, d\phi^2+\frac{2Mr}{\Sigma}(dt-a\sin^2\theta\, d\phi)^2\,,
\end{equation}
where $M$ and $J > 0$ are the mass and angular momentum of the gravitational source, respectively, $a = J/(Mc)$ has dimensions of length and is the specific angular momentum of the source, $(t,r,\theta,\phi)$ are the standard Boyer-Lindquist coordinates and
\begin{equation}\label{E2}
\Sigma=r^2+a^2\cos^2\theta\,,\qquad \Delta=r^2-2Mr+a^2\,.
\end{equation}
The Kerr metric contains gravitoelectric, $GM/(c^2r)$, and gravitomagnetic, $GJ/(c^3r^2)$,  potentials that correspond to the mass and angular momentum of the source, respectively. 

The QNM frequencies of the Kerr metric using the light-ring method have been treated in detail in Ref.~\cite{FeMa2} and can be expressed as
\begin{equation}\label{E3}
\omega_{QNM} = \omega^0_{K} + i \Gamma_{K}  = \pm \,j \omega_{\pm}^K + i\gamma_{K} \left( n + \frac{1}{2} \right)\,, \qquad n = 0, 1, 2, 3, \cdots\,,
\end{equation}
where $\gamma_{K} > 0$ and
\begin{equation}\label{E4}
\omega_{\pm}^K = \pm \frac{\sqrt{M/r_0^3}}{1\pm a \sqrt{M/r_0^3}}\,,\quad   \gamma_{K} = \sqrt{3}\,|\omega_{\pm}^K|\,\left(1-\frac{2M}{r_0} +\frac{a^2}{r_0^2}\right)^{1/2}\,, \quad r_0 = 3M\mp 2a\left(\frac{M}{r_0}\right)^{1/2}\,.
\end{equation}

To second order in the dimensionless parameter $a/M$, the radius of the light ring is given by
\begin{align}\label{E5}
r_0 = 3M \mp \frac{2a}{\sqrt{3}} - \frac{2a^2}{9M}\,,
\end{align}
so that at this level of approximation 
\begin{align}\label{E6}
\omega_{\pm}^K = \pm\frac{1}{3\sqrt{3}M}\,\left(1 \pm \frac{2a}{3\sqrt{3}M} +\frac{11}{54}\, \frac{a^2}{M^2}\right)\,, \qquad \gamma_K = \frac{1}{3\sqrt{3}M}\,\left(1  - \frac{2}{27}\,\frac{a^2}{M^2}\right)\,.
\end{align}
These results agree with the QNM frequencies of the rotating Hartle-Thorne solution for $Q_K= a^2M$.


\end{document}